\documentclass[acmsmall]{acmart}
\usepackage{multirow}
\usepackage{listings}

\usepackage{caption}
\usepackage{graphicx}
\usepackage{longtable}
\usepackage{array}
\usepackage{booktabs}
\usepackage{ragged2e}
\usepackage{subcaption}
 \usepackage{blindtext}
 \usepackage{makecell}
\usepackage{xurl}

\usepackage[T1]{fontenc}

\def\BibTeX{{\rm B\kern-.05em{\sc i\kern-.025em b}\kern-.08emT\kern-.1667em\lower.7ex\hbox{E}\kern-.125emX}}

\setcopyright{none}

\acmYear{2024} \acmVolume{1} \acmNumber{1} \acmArticle{1} \acmMonth{1} 
\acmDOI{10.1145/XXXX}

\acmISBN{123-4567-24-567/08/06}

\acmJournal{TOCE}

\acmYear{2024} \acmVolume{1} \acmNumber{1} \acmArticle{1} \acmMonth{1} 

\copyrightyear{2024}
\acmPrice{0.00}

\begin{document}

\title[Classroom Activities and New Classroom Apps for Enhancing Children's Understanding of...]{Classroom Activities and New Classroom Apps for Enhancing Children's Understanding of Social Media Mechanisms}

\author{Henriikka Vartiainen}
\email{henriikka.vartiainen@uef.fi}
\affiliation{\institution{University of Eastern Finland}}

\author{Nicolas Pope}
\email{nicolas.pope@uef.fi}
\affiliation{\institution{University of Eastern Finland}}

\author{Juho Kahila}
\email{juho.kahila@uef.fi}
\affiliation{\institution{University of Eastern Finland}}

\author{Sonsoles López-Pernas}
\email{sonsoles.lopez@uef.fi}
\affiliation{\institution{University of Eastern Finland}}

\author{Matti Tedre}
\email{matti.tedre@uef.fi}
\affiliation{\institution{University of Eastern Finland}}

\renewcommand{\shortauthors}{Vartiainen et al.}

\newcommand{\AbstractCategory}[1]{%
  \par\addvspace{.5\baselineskip}%
  \noindent\textbf{#1}\quad\ignorespaces
}

\begin{abstract}
Young people are increasingly exposed to adverse effects of data-driven profiling, recommending, and manipulation on social media platforms, most of them without adequate understanding of the mechanisms that drive these platforms. In the context of computing education, educating learners about mechanisms and data practices of social media may improve young learners' data agency, digital literacy, and understanding how their digital services work. A four-hour technology- supported intervention was designed and implemented in 12 schools involving 209 5th and 8th grade learners. Two new classroom apps were developed to support the classroom activities. Using Likert-scale questions borrowed from a data agency questionnaire and open-ended questions that mapped learners' data-driven reasoning on social media phenomena, this article shows significant improvement between pre- and post-tests in learners' data agency and data-driven explanations of social media mechanisms. Results present an example of improving young learners' understanding of social media mechanisms.
\end{abstract}



\maketitle

\section{Introduction}

Children and young people today grow up in a new media ecology dominated
by social media platforms that blend social content, news, opinions,
advertising, and entertainment, often blurring the boundaries between
them (Ofcom, 2023). While most youth are active users of social media
services---and some also active contributors themselves---they often
lack awareness of the various types of personal data these platforms
collect about them, and how these data are used (Pangrazio \& Selwyn,
2019; Authors, 2024). Personal and behavioral data, once considered a
mere byproduct of online participation, have become valuable economic
resources for platform owners, who collect vast amounts of information
about users\textquotesingle{} actions and interactions, such as browsing
habits, locations, emotional reactions, moods, media consumption, and
preferences (McCosker, 2017; van Dijck, 2013; Zuboff, 2015). Such
multidimensional data, shared voluntarily or unknowingly, enables the
creation of detailed user profiles at a global scale. Those data, in
turn, enable platforms to deliver targeted and personalized content
aimed at maximizing user engagement on the platforms (Fisher, 2022).

Algorithmic processes increasingly influence people's decisions and
actions, categorizing, classifying, and prioritizing people, places,
objects, ideas, and even their thought patterns (Hallinan and Striphas,
2016). In this algorithmic culture, people's personal choices are
increasingly influenced by their social networks and by like-minded
individuals who have made similar decisions in comparable situations
(van Dijck, 2013). These data-driven practices of influencing enable and
create interactions, emotional responses, behaviors, and new habits of
mind that might not exist without large-scale data collection and use
(Hallinan \& Striphas, 2016; Vartiainen \& Tedre, 2024). Furthermore,
tracking and profiling facilitate targeted marketing, behavior
engineering, attention harvesting, and spread of targeted or tailored
mis- and disinformation, as well as many other less desirable
data-driven practices (Hendricks \& Vestergaard, 2018; Kramer et al.,
2014; Valtonen et al., 2019).

A critical discourse is growing in the context of the complexity of
human agency in online environments where people's interactions,
decision-making, knowledge construction, and participation are
increasingly mediated, augmented, produced, and regulated by AI-driven
systems (Kitchin, 2012; Vartiainen \& Tedre, 2024). In light of these
transformations, it is essential for online users to understand the
mechanisms that drive social media (Pangrazio \& Selwyn, 2019). This
understanding enables them to make informed, ethical, and critical
decisions that are grounded in understanding how data are generated,
processed, and used for different purposes (Authors, xxx). What makes
that understanding difficult is that those mechanisms are naturally
opaque to some extent, as well as protected secrets of platform owners.
One challenge of media education in the 2020s is how to expose those
mechanisms and enhance young learners' understanding of social media
mechanisms and data practices that they encounter daily (Grover, 2024;
Valtonen et al., 2019; Morales-Navarro et al., 2024; Heintz \& Roos,
2021; Long \& Magerko, 2020; Shapiro \& Fiebrink, 2019; Chiu et al.,
2024).

Despite the widespread use of social media and other online platforms in
young people's lives, understanding AI and other mechanisms of social
media are often missing in educational curricula (e.g., Van Mechelen et
al., 2023; Martins \& Gresse Von Wangenheim, 2022). This risks increased
reliance on folk theories (DeVito et al., 2017, 2018), new digital
divides (Grover, 2024), and overlooking significant ethical questions
(Morales-Navarro et al., 2024; Druga et al., 2022). There is a marked
lack of research on how those mechanisms can be taught to young learners
in a media-rich environment (for recent approaches see, e.g.,
Morales-Navarro et al., 2024; Van Mechelen et al., 2023; Rizvi et al.,
2023). This study presents findings from research conducted in 12
classrooms in Finnish elementary and secondary schools (classes 5 and 8,
11--15 years old children). This article presents hands-on,
computer-supported classroom activities and educational technologies
designed to facilitate collaborative learning and critical inquiry. It
presents results from engaging 209 students in exploring how social
media works, including its basic concepts and principles, as well as
exposing some of its opaque mechanisms. The study posed two research
questions:

\begin{enumerate}
\def\labelenumi{\arabic{enumi}.}
\item
  How do children's explanations of the basic mechanisms of social
  media develop over a four-hour technology-rich educational
  intervention?\\
\item
  How do children's perceptions of their data agency in relation to
  social media develop over that intervention?
\end{enumerate}

This article is aimed at computing education researchers, educators,
policymakers, researchers, and educational technology researchers
interested in enhancing data agency and social media literacy among
young people (Vartiainen et al., 2024; for data awareness, see Höper \&
Schulte, 2024). By providing practical insights and evidence-based
strategies for the classroom, we hope to contribute to computing
education research that focuses on data-driven practices of social media
and strengthened agency in the datafied society.\\

This article begins by introducing the key concepts of social media that
children explored in the intervention, namely, data collection,
profiling, engagement, and recommendation. Then, it takes a look at the
previous research on people\textquotesingle s everyday conceptions and
folk theories concerning social media, and how they are connected to
pedagogical challenges of fostering the development of understanding of
scientific concepts that differ from everyday experiences and
observations. The article details the study context, the instructional
intervention, and the educational technologies and learning activities
designed to cultivate collaborative learning and inquiry. After
describing data collection and analysis, and results of pre-post tests,
the article ends with a discussion on the possibilities and challenges
of supporting children\textquotesingle s understanding of social media
mechanisms, and how this should be taken into account when facilitating
informative and agentive actions in the age of AI.

\section{Unpacking some basic concepts}
Computer-based processing of personal data has vexed people since
the advent of modern computing (Acquisti, 2015). As society's functions
have become increasingly computerized, theorists have explored how
information systems influence various aspects of life, including work
dynamics (Zuboff, 1988), worldviews (Bowker \& Star, 2000), social
connections (boyd \& Ellison, 2007), and identity formation (boyd,
2015), among many others. The 2000s saw the rise of pervasive computing,
the Internet of Things, data-intensive analysis, and social media, which
fueled unprecedented capabilities for tracking and profiling
individuals, providing detailed insights into personal behaviors (Swan,
2012; Zuboff, 2019). Extensive interdisciplinary body of research has
accrued on the gradual loss of privacy from perspectives like rights,
commodities, and control (Smith et al., 2011; Acquisti, 2015)---and as a
response, it has been suggested that everyone should learn and
understand how massive amounts of data, including location, moods,
preferences, and personal secrets, are collected and processed by
everyday online platforms (Bowler et al., 2017; Swan, 2012; Valtonen et
al., 2019).

Understanding the basics of datafication and the data-driven
mechanisms of social media require familiarity with a number of
concepts, of which four were selected for this intervention. The first
is \emph{data collection} and different types of data involved (e.g.,
van der Hof, 2016; Pangrazio \& Selwyn, 2019). As users sign up they
provide volunteered data, or ``data given'', such as name, age, and
location. As they interact with the platform, their actions provide
observed data, or ``data traces'', such as mouse clicks, time spent on
content, keyboard typing speed, or software accessibility settings
(Jünger, 2018). And once enough directly observed data has accumulated,
the systems can statistically or otherwise infer missing data about
users by comparing them with similar users (``data inferred'')---data
such as movie preferences, spending habits, and political leanings
(Hendricks \& Vestergaard, 2019). The volume, velocity, and variety of
data is key to understanding its privacy and security implications
(e.g., Barassi, 2020)---but the extent of data collection is opaque to
users.

The second key concept is \emph{profiling}, which involves categorizing
individuals into innumerably many categories based on their behaviors,
interests, and demographics, among other things (Bowker \& Star, 2000).
This process is messy and complex, inherently fuzzy, discriminating by
default, and invariably inaccurate (Bowker \& Star, 2000; Noble, 2018;
Benjamin, 2019; Crawford, 2021; Morales-Navarro et al., 2024). Despite
its shortcomings, profiling underpins effective online experiences, such
as personalized interfaces, tailored content, and precise
recommendations---but it is also crucial for targeted advertising,
opinion swaying, behavior engineering, and directed manipulation
(Valtonen et al., 2019; Eubanks, 2018; Zuboff, 2019). The EU GDPR
mandates that users should be able to download and inspect their
personal data, including their profiles, but in practice the data are
too complex and too voluminous to be human-readable.

The third key concept is \emph{engagement}, or the user's interaction
with content on a social media platform, involving both data given
explicitly (e.g,. liking or commenting) and unobtrusively collected data
traces (e.g., stopping to view content instead of scrolling through)
(Dolan et al., 2016). The social media business model often relies on
maximizing user engagement to increase user retention and attract
advertisers, at the cost of some users' well-being (Valkenburg et al.,
2022). Social media sites calculate engagement scores based on data
given, data traces, and inferred data, but those scores and their
calculation methods remain invisible to the user.

The fourth key concept is \emph{recommending}, which is central to
maximizing user engagement by suggesting, for example, content, friends,
and communities that align with the user's interests. Recommending
relies on the profile built of each user, the profiles of similar users,
and a range of techniques such as collaborative filtering (the more one
user's interests align with another on certain topics, the higher the
likelihood that their interests will align on other topics as well).
Recommending is typically overt, but its actual mechanisms are opaque to
the users.

While the data-driven mechanisms of social media are hidden from the
users, the epistemic opacity behind these concepts and practices also
poses novel theoretical and practical challenges for education. That
opacity comes in many forms; some opacity is essential (e.g., intrinsic
complexity of the mechanisms) and some opacity is accidental (e.g.,
contingency or human choices) (Koskinen, 2023). Despite extensive
research on promoting conceptual and scientific understanding in school
education (Aleknavičiūtė et al., 2023; Lehtinen et al., 2020), social
media mechanisms differ significantly from the well-defined concepts and
transparent mechanisms in many other disciplines. Many AI systems lack
precise and complete scientific representations due to a variety of
intrinsic and contingent opaque features.
\section{Fostering conceptual change}

While the rise of social media has not been overlooked in
educational research, most approaches to media literacy have
predominantly focused on children's and young people's media use,
their practices, cultures of creative participation, and their
ability to interpret, evaluate, and critique online content (e.g., Kafai et al., 2018; Livingstone et al., 2019; Rasi et al., 2019;
Morales-Navarro et al., 2024). Media education has also emphasized the
importance of encouraging youth to engage in critical and ethical
reflections on the nature of new media (Kafai et al., 2018;
Morales-Navarro et al., 2024) and understanding the intentions,
applications, and purposes behind social media platforms (Valtonen et
al., 2019). However, topics related to the underlying mechanisms of
social media and the datafication of everyday life are often missing
from educational curricula (Pangrazio \& Selwyn, 2019; Stoilova et al.,
2020; Van Mechelen et al., 2023). Moreover, there are substantial
disparities in the support children and youth receive from their parents
and teachers concerning the data economy and digital infrastructure
underpinning everyday interactions (Stoilova et al., 2020).

Previous research has shown that people intuitively develop various
kinds of folk theories to explain AI or social media systems and
mechanisms (DeVito et al., 2018; Mühling \& Große-Bölting, 2023). DeVito
et al. (2017) defined folk theories as \emph{``intuitive, informal
theories that individuals develop to explain the outcomes, effects, or
consequences of technological systems, which guide reactions to and
behavior towards said systems.''}  These theories encompass causal
models of how algorithms work, as well as opinions and attitudes
 regarding potential outcomes of their operations (DeVito et al.,
2017). These intuitive theories are generated through everyday use of
social media as well as through other sources such as media or
discussions with friends (DeVito et al., 2018). Folk theories may
significantly differ from scientific theories or engineering
explanations, as they require users to generalize the inner workings of
algorithms, which are hidden behind simple interfaces (Büchi
et al., 2023). Yet, folk theories essentially act as the contextual
frames through which people interact with social media and form
assumptions and expectations that guide their actions, reactions,
identity-formation, and understanding (DeVito et al., 2017; Karizat et
al., 2021). Recent research also indicates that the formation of more
scientific conceptions of AI and related mechanisms is unlikely to
happen through everyday experiences and interactions with AI among
primary school students (Mertala et al., 2022; Authors, 2024) and
preservice teachers (Vartiainen et al., 2024).

While folk theories about the inner workings of social media have
primarily been studied in the field of human-computer interaction,
educational research on children\textquotesingle s conceptions and
conceptual change has extensively studied the difficulties students face
in formal educational settings when attempting to develop their
understanding of certain scientific concepts (Aleknavičiūtė et al.,
2023; Lehtinen et al., 2020). Conceptual change research has revealed
that children construct various kinds of naïve conceptions or
misconceptions from their everyday experiences and observations from lay
culture (Chi, 2008; Vosniadou et al., 2008). Thus, children already
possess initial conceptions and prior knowledge that not only shape how
they interpret the world around them, but very often, these folk
theories can incorporate misconceptions that have contradictory impacts
on learning in classrooms (Aleknavičiūtė, Lehtinen \& Södervik, 2023).

Studies on conceptual change have also shown that naïve conceptions and
misinterpretations are difficult to change, as initial conceptions
may persist even after many years of exposure to instruction (Vosniadou,
2013). Conceptual change is a long and gradual process during which new
information is incorporated into the existing knowledge base
(Vosniadou, 2013). However, in many cases, learning advanced
scientific concepts does not lead to the replacement of initial
concepts; rather, the scientific and initial concepts can coexist
(Vosniadou, 2014). In other words, slow revision of an initial
conceptual system may lead towards scientific understanding, but also result in various misconceptions or synthetic conceptions that integrate some aspects of the scientific concepts with initial conceptions and beliefs (Vosniadou, 2014).

Research on conceptual change has suggested that understanding
scientific phenomena can and should be facilitated (Aleknavičiūtė et
al., 2023; Schlatter et al., 2020). Advances and revisions in children's
explanations can be supported by hands-on learning experiences where
children are guided to test their prior conceptions by conducting
experiments and explaining their conclusions with systematic evidence
(Aleknavičiūtė et al., 2023; Osterhaus et al., 2021). Previous research
has also demonstrated that computer simulations and games can support
the development of understanding by making abstract scientific concepts
more accessible, concrete, and visible (Smetana \& Bell, 2012; Trundle
\& Bell, 2010). They allow students to confront their own beliefs by
visualizing objects and processes that are normally beyond perception,
and by enabling students to manipulate variables that are otherwise
beyond their control in the physical world (Trundle \& Bell, 2010;
Smetana \& Bell, 2012).

Individual learning and conceptual change can also be facilitated
through collaborative activities that cultivate new ways of thinking
through joint exploration, dialogical interaction, argumentation, and
classroom discussions (Vosniadou, 2007). Students' prior conceptions and
modes of thought should also be made visible to the teacher, who can
then design, tailor, and scaffold the construction of scientific
understanding so that students experience dissatisfaction with their
prior conceptions and are motivated to restructure their prior knowledge
and understandings (Aleknavičiūtė et al., 2023).

Although research on how to support children\textquotesingle s
understanding and reasoning of the underlying mechanisms of social media
and AI is scarce, some studies suggest that collaborative learning and
exploration of ML-based technologies and mechanisms could offer a
promising avenue for students to develop their conceptual understanding
of ML principles, workflows, and its applications (e.g., Grover, 2024).
Yet, to our knowledge, no studies have examined changes in
children\textquotesingle s conceptions of social media mechanisms as a
result of collaborative inquiry using computer simulations, games, and
learning activities designed to expose the processes by which familiar
social media platforms track users, profile them, and recommend content
tailored to individual preferences.

\section{Methodology}

\subsection{Context and study design}

This research is part of a long-term, multidisciplinary design-based
research and a larger research program focused on co-designing
educational technologies, pedagogical models, and scaffolding strategies
for integrating AI topics to school education (Authors, 2024a; 2024b).
In the program, initiated in 2023, researchers, software developers, and
schoolteachers co-design and implement a series of increasingly complex
and cumulative projects in local schools, involving over 200 school
children. This paper presents findings from a second round of
interventions conducted in spring 2024 with the same classrooms that
participated in the first round in spring 2023.

\subsection{Participants}

The participants of the study were twelve classes of 5th and 8th graders
(N=209) from Finnish elementary and secondary schools (Table 1).

\begin{table}[ht]
 \centering
  \caption{Number of participants per grade}

  \begin{tabular}{@{} >{\raggedright\arraybackslash}p{0.307\columnwidth} 
                         >{\raggedright\arraybackslash}p{0.2608\columnwidth} 
                         >{\raggedright\arraybackslash}p{0.3885\columnwidth} @{}}
    \toprule
    & \textbf{Participants} & \textbf{Schools (classrooms)} \\
    \midrule
    5th grade & 129 & 7 \\
    7th grade & 80 & 5 \\
    \textbf{Total} & 209 & 12 \\
    \bottomrule
  \end{tabular}
  \label{your-label}
\end{table}

The study was conducted following the guidelines of the Finnish National
Board on Research Integrity (TENK, 2019). Research permit was obtained
from the municipal educational administration, and informed consent was
also obtained from the guardians of each participant. In addition,
researchers verbally informed the children about the research
methodology and data collection process, emphasizing the confidential
and voluntary nature of the study. As the workshops were integrated into
the regular school activities and implemented during school hours, the
workshops involved participating and non-participating students, and
only data from the former were collected.

\subsubsection{Instructional intervention}

The school projects involved two workshops, each lasting approximately
1.5 to 2 hours. These workshops were led by a researcher experienced in
primary school classroom teaching who had previously worked with the
same students during the first-year school interventions in 2023. In
these two workshops, the children explored social media mechanisms using
two newly developed educational classroom apps designed to facilitate
collaborative learning and inquiry.

\paragraph{The first workshop}
The first workshop focused on introducing students to key concepts in
datafication and social media---data collection, profiling, engagement,
and recommendation---through gamified learning activities centered
around a profiling game (for more detailed description of the
educational app, see Authors, 2024a). In this game, students were guided
to analyze various types of everyday data traces to construct and
iteratively refine detailed profiles of a mystery social media user.
These data traces were similar to those collected by social media
services commonly used by children and youth (Authors, forthcoming). The
game provided relatable examples for children to investigate, such as
listening to music, watching videos, following someone, posting images,
creating stories, searching for information, and sharing locations.
Storylines featuring familiar characters and their daily activities
helped connect the datafication-related concepts to the students'
everyday lives.

\begin{figure}
    \centering
    \includegraphics[width=4.11111in,height=3.07992in]{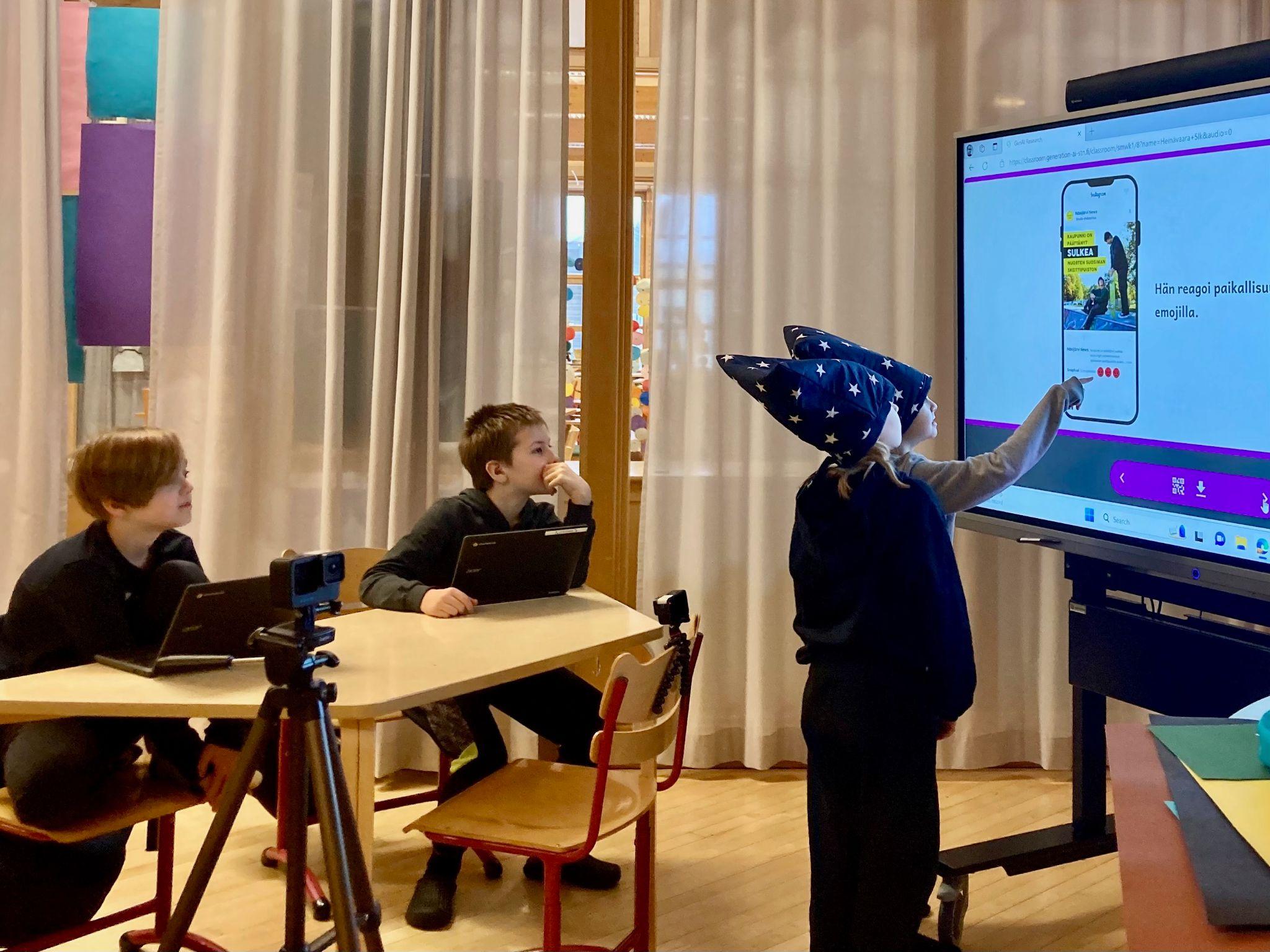}
    \caption{Children combining data traces to form a profile of the
mystery person X. The teacher's screen shows one hint at a time, and
each student pair builds and gradually refines a profile of X on their
personal devices.}
    \label{fig:fig1}
\end{figure}

The activity was designed to demonstrate how everyday actions on social
media and other online platforms generate data that can be used to infer
and describe one\textquotesingle s persona, preferences, moods, and
behaviors. It also introduced the concept of profiling, showing how data
from different profiles can be combined to create detailed profiles and
predict interests. The activity highlighted the differences between
human-made and automated profiling techniques, which were further
elaborated in the second workshop. Figure \ref{fig:fig1} presents a picture of the
classroom activity.

To foster collaborative learning and inquiry, children were paired up to
work together during the game. The game was designed to support
collaborative discussions by gradually unveiling new clues and by
presenting question prompts. These prompts encouraged students to
externalize, articulate, and reflect on their reasoning and make
connections across different data traces. The students were scaffolded
to collaboratively construct and reconstruct their evolving
understanding by refining and evaluating the profiles which they were
building. The sociotechnical design of the game further supports
collaboration in that the game concludes with profiles built by student
pairs shared and projected on the classroom screen, using the game's
virtual bulletin board feature. This allowed all students to see how
others had profiled the same mystery person and engaged the whole class
in discussions facilitated by the teacher.

After the game activity, the teacher instructed students to reflect on
their learning, discuss in pairs, and write down the types of data that
can be collected from their own social media activities. Students were
prompted to think about what can be inferred from those data (similar to
what they inferred in the game), and to consider the purposes for which
these inferred data might be used. By linking the game activities to
their own everyday experiences, the aim was to scaffold children to
reflect and reconsider their own social media activities and data traces
using datafication-related concepts and computing's disciplinary
strategies.

\paragraph{The second workshop} 

\begin{figure}
    \centering
    \begin{subfigure}[b]{0.475\textwidth}
        \centering
\includegraphics[height=3.8in]{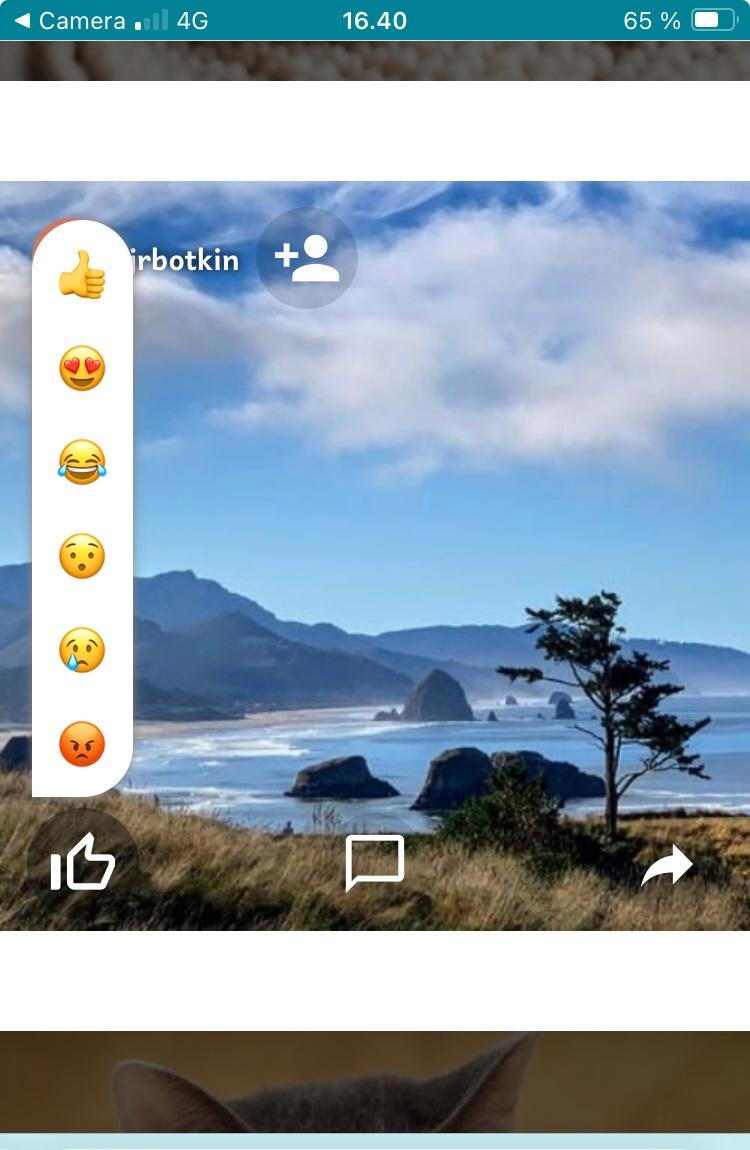}
    \end{subfigure}
    \hfill
    \begin{subfigure}[b]{0.475\textwidth}
        \centering
\includegraphics[height=3.8in]{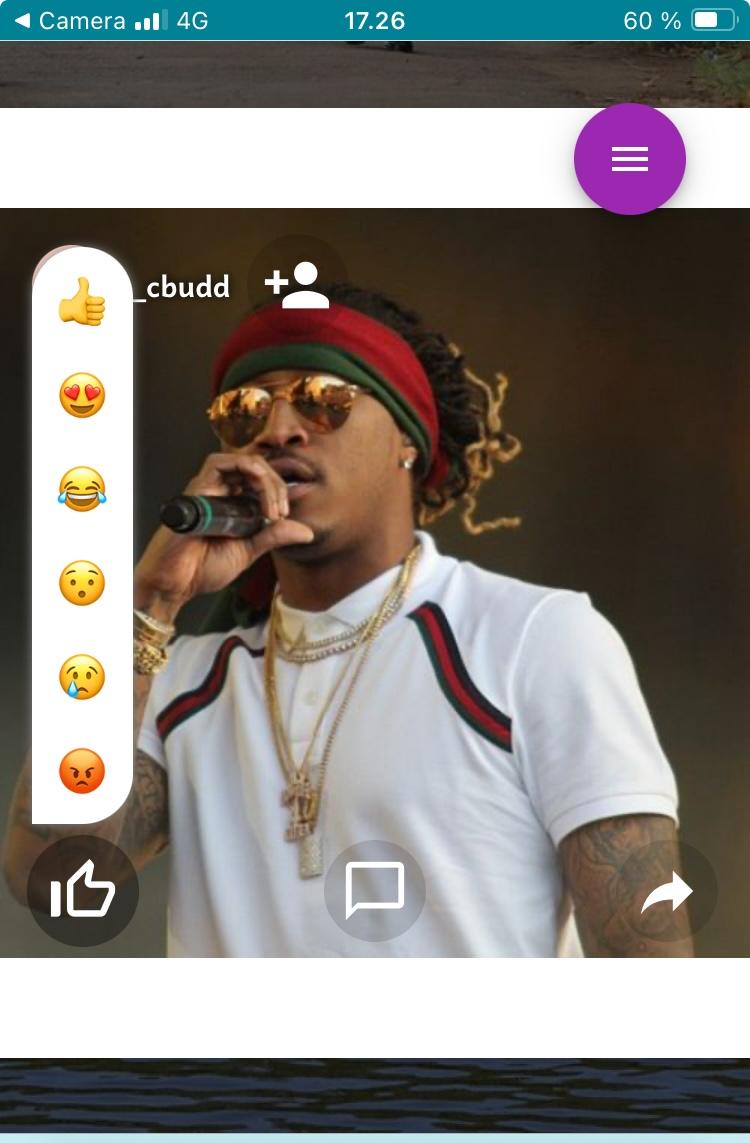}
    \end{subfigure}
 
    \caption{Screenshots of user browsing images on Somekone, with ``like'' button selected to react to pictures.}%
    \label{fig:fig2}%
\end{figure}

The second workshop was intended to elaborate and deepen the exploration
of mechanisms and concepts related to tracking, profiling, engagement,
and content recommendation. To support children\textquotesingle s
collaborative inquiry, a new educational classroom app called
''Somekone'' was developed (for more detailed description of the app,
see Authors, 2024b). Resembling Instagram's user interface, Somekone
offers an infinite feed of images along with typical social media
engagement features like liking, reacting, commenting, following, and
sharing (Fig. \ref{fig:fig2}). It also provides the ability to pair one's device
with another device that presents a real-time visualization of how the
social media platform Somekone tracks users, profiles them, and tailors
recommendations.

Students worked in pairs, with one device used to browse the Somekone
feed and another device as an analytics platform. Students were guided
to collaboratively explore the data generated by their browsing, the
profiles being built from them, and the upcoming recommendations formed
from engagement data collected from all users in the classroom. Children
were instructed to imagine browsing as someone else, to use a nickname
instead of their real name, and avoid sharing any sensitive information,
as everyone's activities would be shown on the classroom screen. The
image dataset was extracted from a classroom safe image board, and then
curated by teachers and researchers (Authors, 2024b).

\begin{figure}
    \centering
    \begin{subfigure}[b]{0.15\textwidth}
        \centering
\includegraphics[height=2.5in]{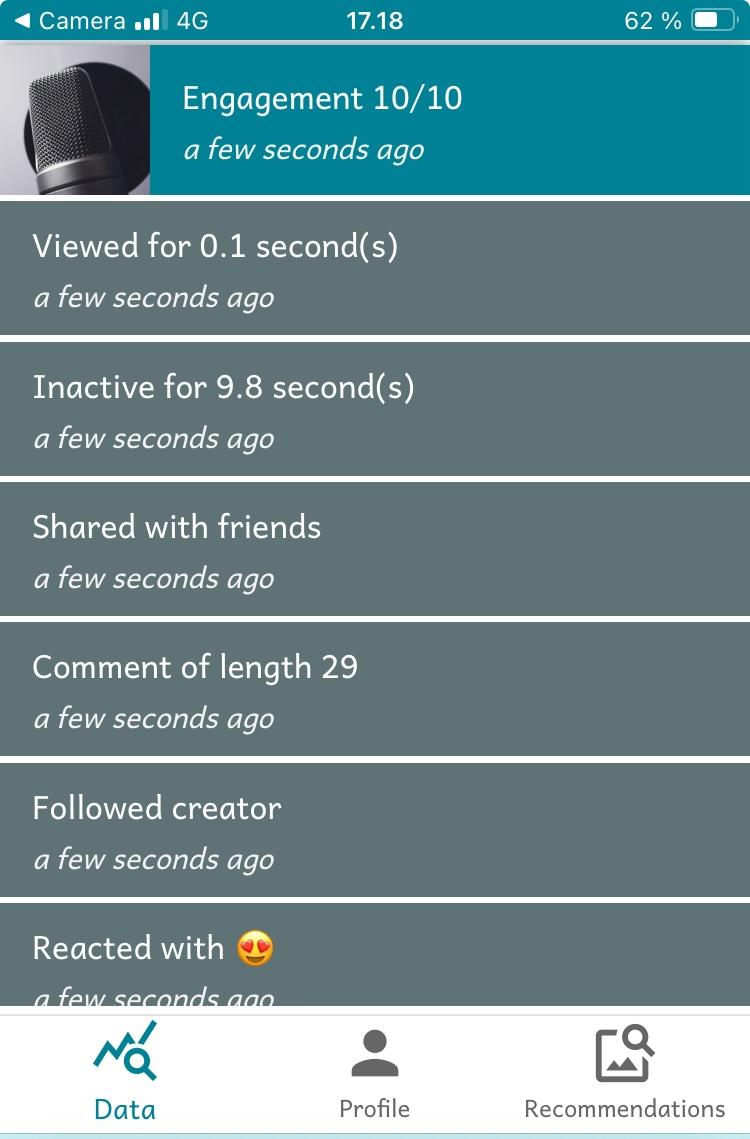}
    \end{subfigure}
    \hfill   
    \begin{subfigure}[b]{0.65\textwidth}
        \centering
\includegraphics[height=2.5in]{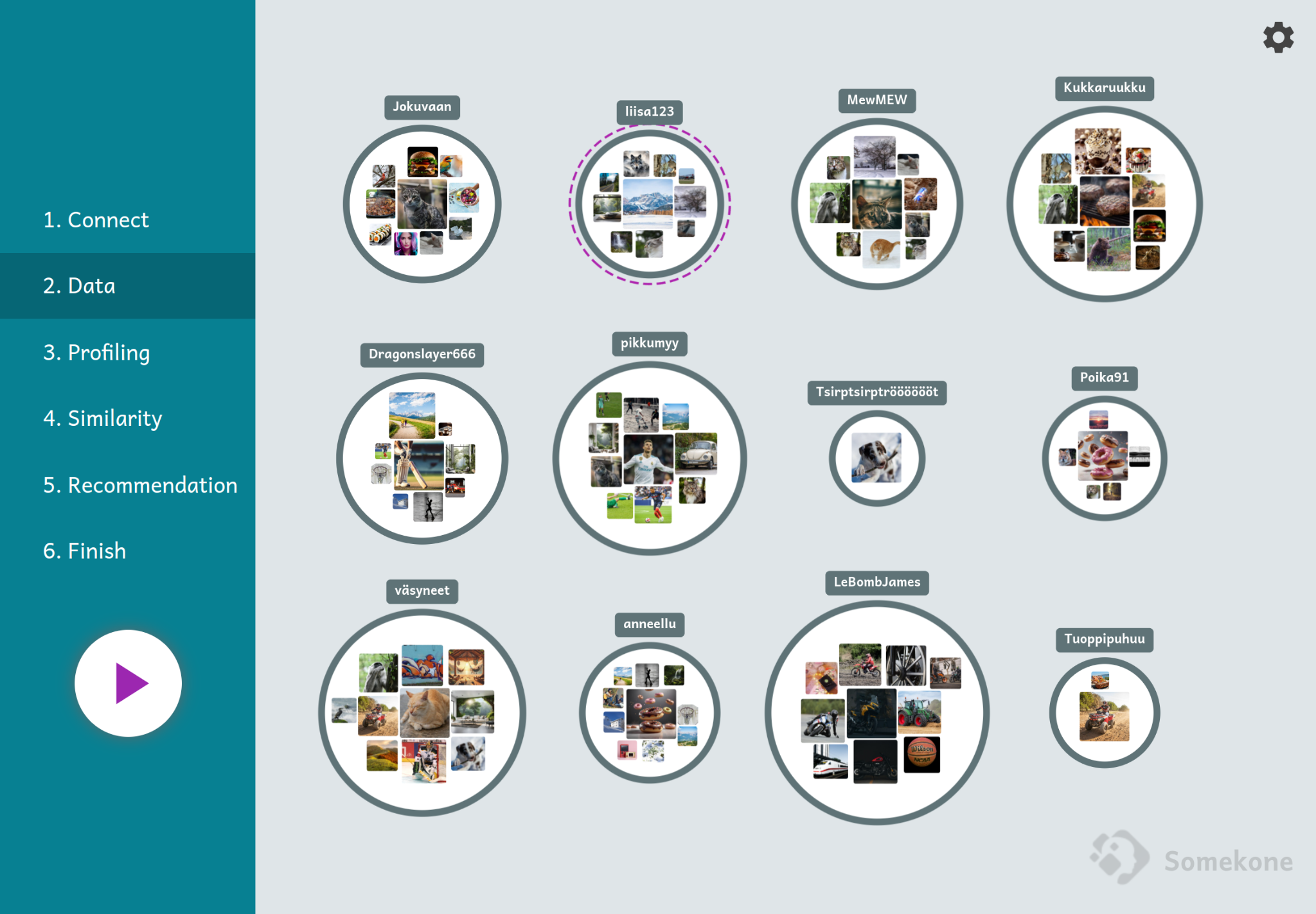}
    \end{subfigure}
    \caption{a) Live view of data collected from another student's
browsing session on a paired device (on the left) and b) the whole
classroom's profiles (on the right).}%
    \label{fig:fig3}%
\end{figure}

Firstly, the concept of tracking was illustrated by providing students
with access to a real-time visualization of the user's action log during
their browsing sessions (Fig.  \ref{fig:fig3}a). As children browsed the feed, they
could explore various types of data traces constantly collected from
their actions on the platform and experiment with how different data
traces affected the engagement score calculated for each image. The data
traces included image views, time spent viewing, likes, comments,
periods of user inactivity, reactions with different emojis, following
other users, and sharing content privately, with friends, or publicly
(see Authors, 2024b). The teacher's projector also showed everyone's
profiles forming up (Fig. \ref{fig:fig3}b).

\begin{figure}
    \centering
    \begin{subfigure}[b]{0.15\textwidth}
        \centering
\includegraphics[height=2.5in]{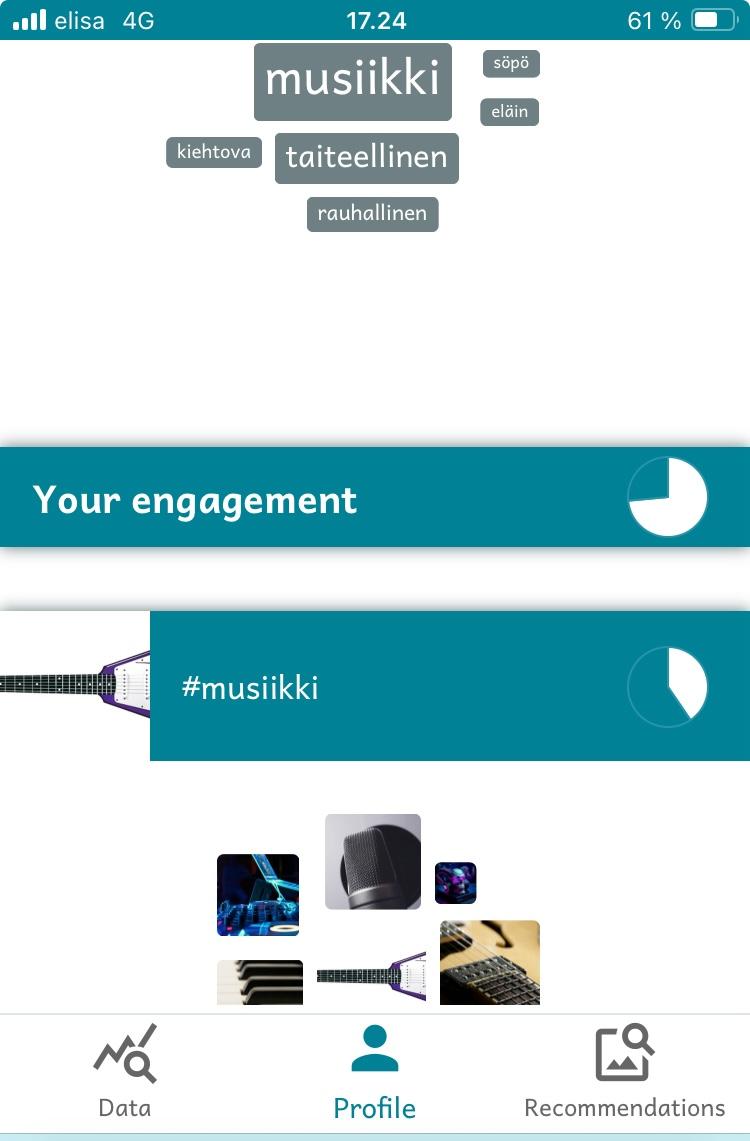}
    \end{subfigure}
    \hfill  
    \begin{subfigure}[b]{0.65\textwidth}
        \centering
\includegraphics[height=2.5in]{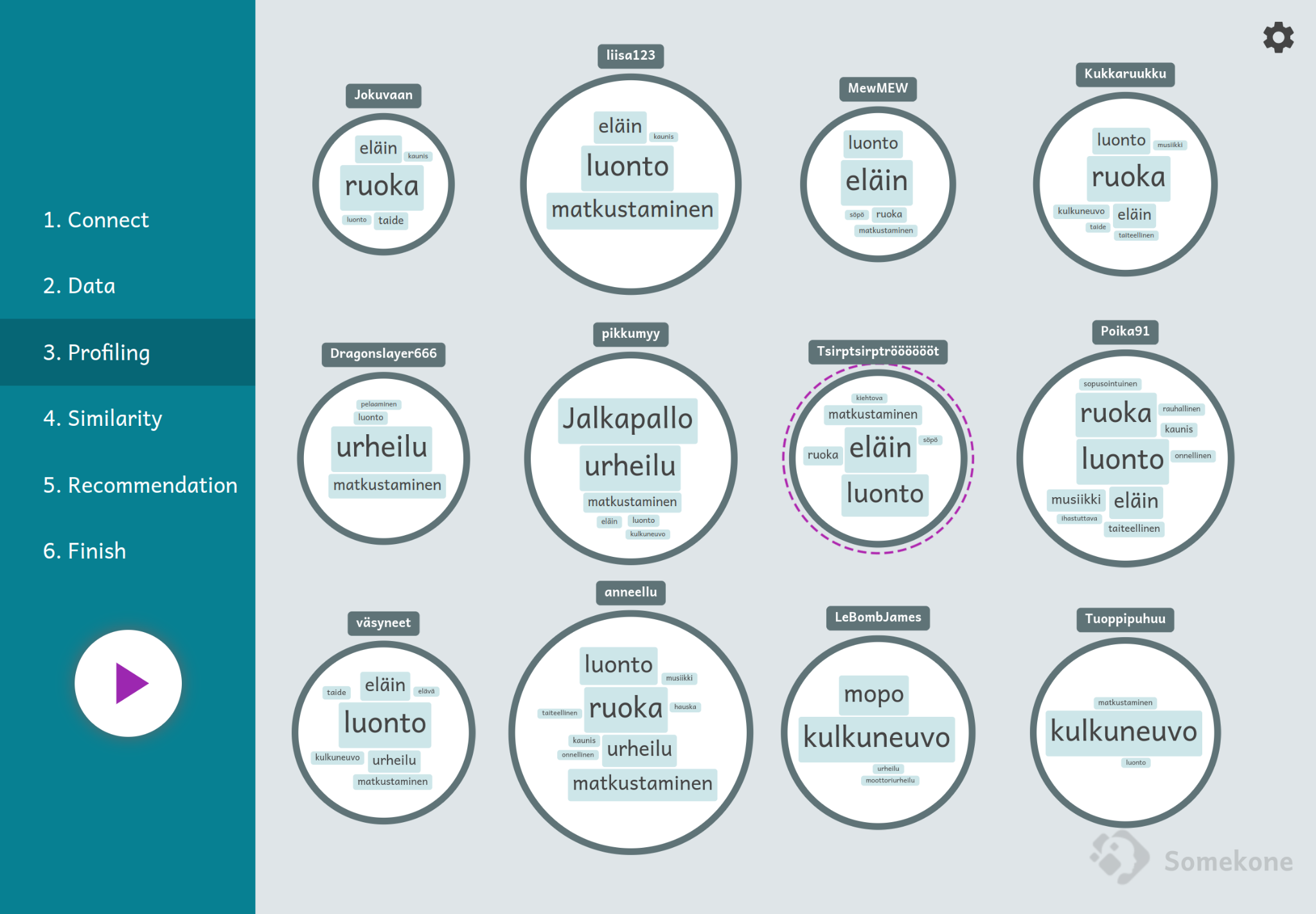}
    \end{subfigure}
    \caption{a) Somekone-generated live view of profile forming up on
a paired device (on the left) and b) the whole classroom of profiles
forming up, shown on the classroom projector (on the right).}%
    \label{fig:fig4}%
\end{figure}

Secondly, the concept of profiling was demonstrated by showing how data
traces are automatically converted into profiles. These profiles
primarily consisted of topic affinities that were continuously updated
based on the children\textquotesingle s own actions and engagement with
labeled images. Figure  \ref{fig:fig4}a shows a screen shot of real-time visualization
on a student's mobile phone. Through these visualizations, the children
could explore how each type of interaction contributed differently to
the engagement score and identify which topics received the most
engagement through actions like sharing, following, reacting with
emojis, and viewing. After exploring how engagement scores were
calculated, the students could then observe how the tool builds profiles
of them, visualized as a word cloud representing their most engaged
topics. The gradually forming profiles from the whole class were
projected on the teacher's screen to ease comparison. Figure \ref{fig:fig4}b shows
the classroom view of all profiles forming up. The students were guided
to evaluate the accuracy of their profiles and compare different
profiles on the teacher's screen.

\begin{figure}
    \centering
    \begin{subfigure}[b]{0.15\textwidth}
        \centering
\includegraphics[height=2.5in ]{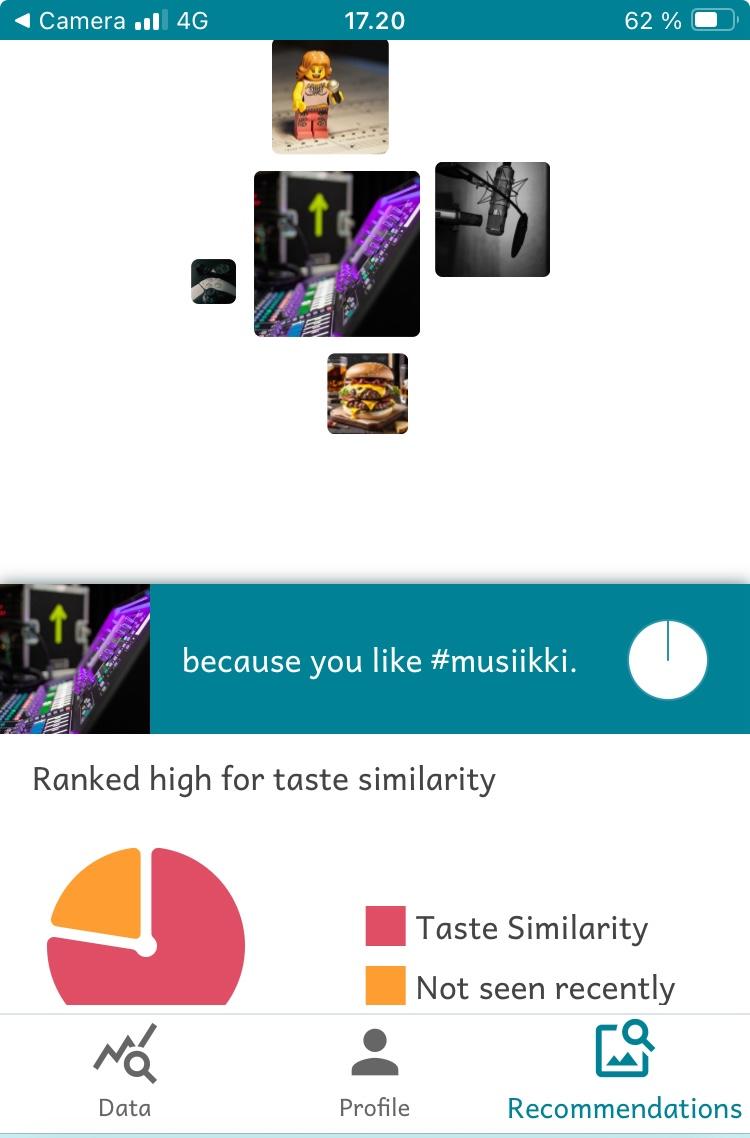}
    \end{subfigure}
    \hfill  
    \begin{subfigure}[b]{0.66\textwidth}
        \centering
\includegraphics[height=2.5in ]{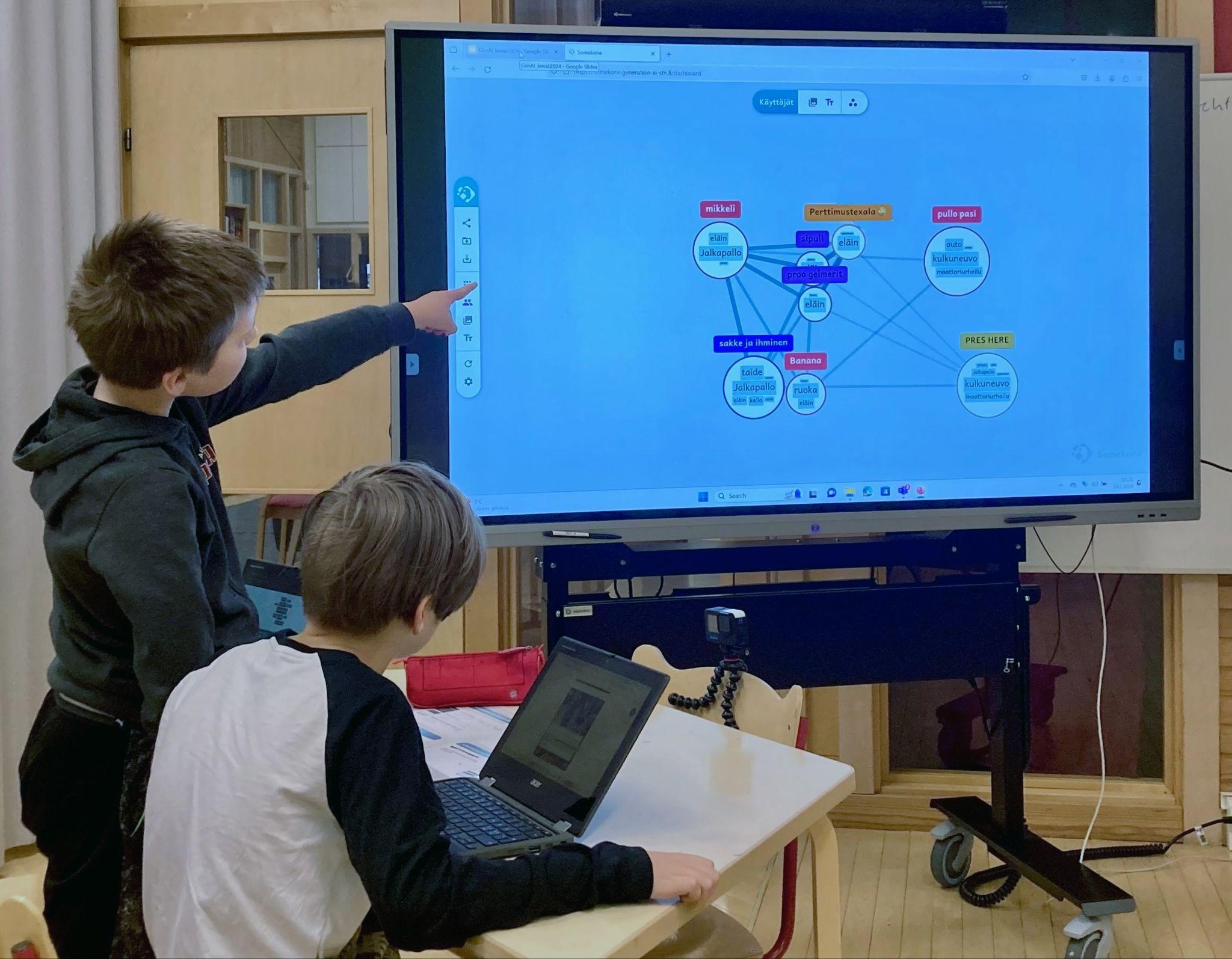}    \end{subfigure}
    \caption{a) A live feed of the five next upcoming recommendations shown on the app (on the left) and b) a real-time visualization of social network of the classroom, clustered by profile similarity (on the right).}%
    \label{fig:fig5}%
\end{figure}

Thirdly, the concept of recommending was illustrated by revealing how
profiling and engagement data from all users influenced the selection of
upcoming images on their feed (Fig. \ref{fig:fig5}a). Through this explainable AI-type
visualization, children could see how their past and ongoing
interactions with the tool not only shaped their personal experience but
was also influenced by the decisions and preferences of other users with
the platform. Somekone also provides a real-time visualization of the
social network of the current class, clustered by profile similarity,
which was shown on the teacher's projector (Figure \ref{fig:fig5}b). This allowed
children to collaboratively explore how clusters of similar profiles
were formed, influencing the recommendations they received (Authors,
2024b).

\begin{figure}
    \centering
    \begin{subfigure}[b]{0.24\textwidth}
        \centering
        \frame{\includegraphics[height=1.45in ]{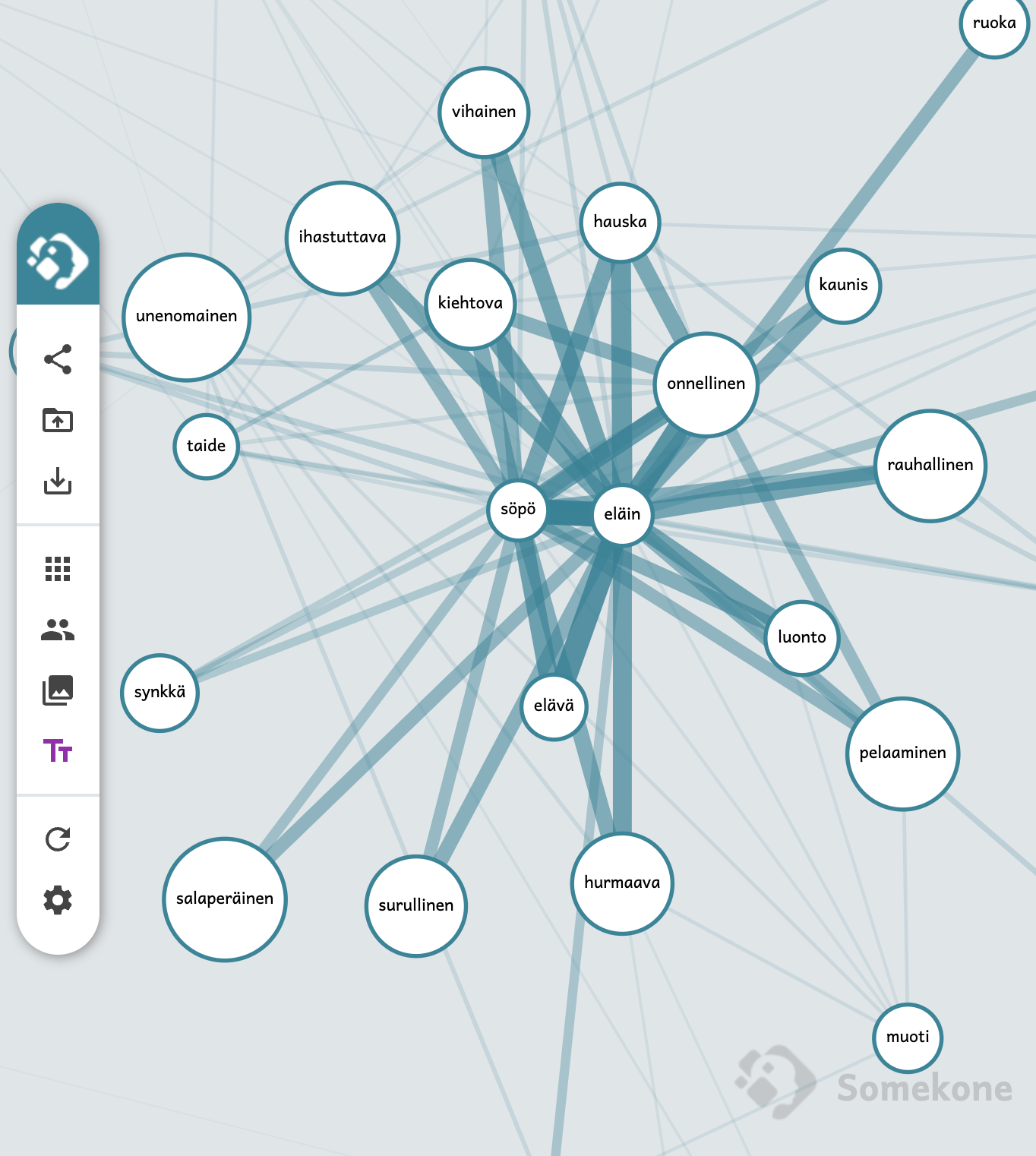}}
    \end{subfigure}
    \hfill  
    \begin{subfigure}[b]{0.24\textwidth}
        \centering
    \frame{\includegraphics[height=1.45in ]{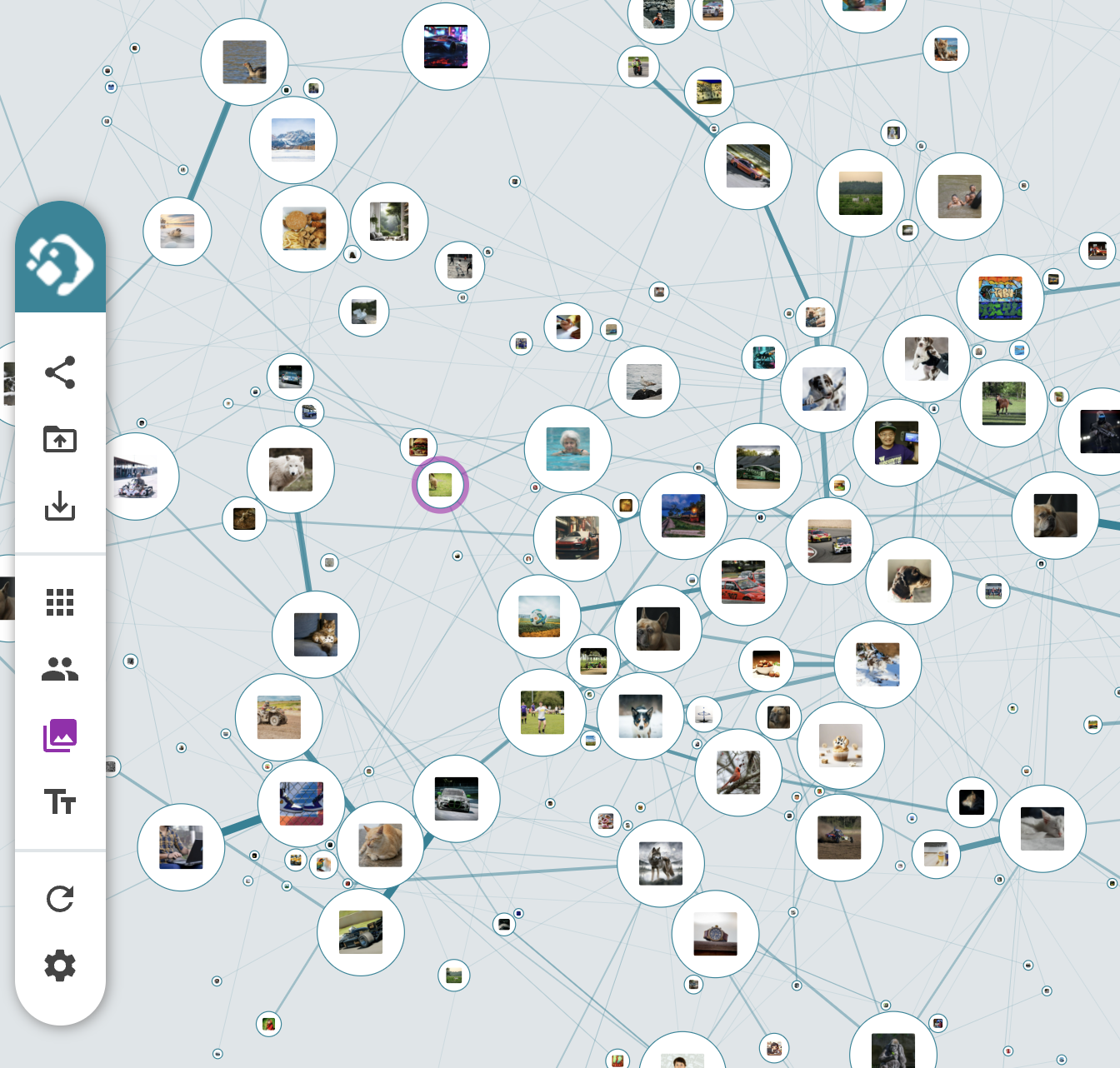}    }   \end{subfigure}
    \begin{subfigure}[b]{0.49\textwidth}
        \centering
    \frame{\includegraphics[height=1.45in ]{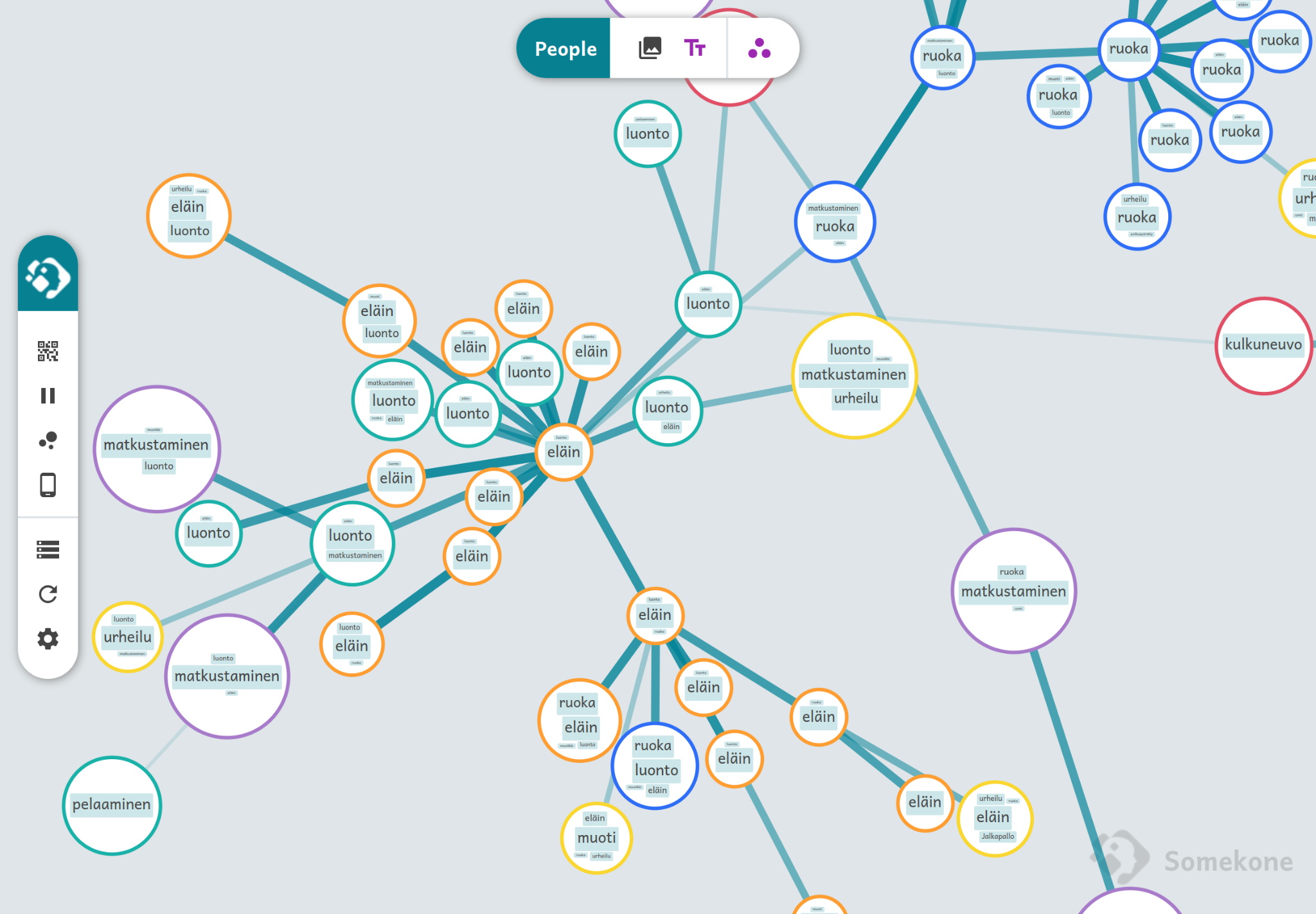}}       \end{subfigure}
    \caption{a) Additional views of the data provided by Somekone: Topic affinities (on the left), b) picture co-engagement network (in the middle), and c) a social network of a large classroom of users (on the right).}%
    \label{fig:fig6}%
\end{figure}

Somekone allows students to collaboratively interact with the system and
explore central social media concepts through their own real-time
actions, facilitated by multiple representations on multiple levels (see
Figure \ref{fig:fig6} for more examples of visualizations presentable for different
age groups). Concepts related to social media mechanisms were anchored
in both individual and collective classroom actions, enabling students
to observe how their micro-level interactions---collected as individual
data traces---lead to visible patterns and clusters at the macro-level,
influencing their recommendations and future interactions.

After the Somekone activity, the teacher gave a presentation about the
impacts of commercial tracking, profiling, and content recommendation.
Children were guided to reflect on the possibilities, potential risks,
and harms that these data-driven practices may pose on oneself and to
others. As a final reflective task, the children wrote a letter to
decision makers about their views on commercial tracking, profiling, and
content recommendation. They were encouraged to propose what might be
done to address their insights and needs. The activity aimed to promote
critical consciousness (Freire, 1974) towards the technologies and power
structures that shape children's everyday lives and to engage children
in considering, suggesting, and imagining alternative possibilities for
change and improvement.

\subsection{Data collection and analysis}
While various kinds of data were collected from the workshop, this study
focused on analyzing pre-test and post-test data. Before the first
workshop, the students were asked to answer a pre-test implemented on a
web form. This online questionnaire was divided into three sections.
The first section measured the use of social software applications (not reported here).

The second section contained 11 statements measuring experiences of
data agency, derived partially from a self-evaluation instrument
developed by Authors (forthcoming). Moreover, it included deliberately
planned items that systematically focused on the understanding of the
target concepts and social media mechanism that the children were
exploring in the interventions (statements are found in Figure  \ref{fig:likert}).
Students had to indicate their level of agreement with the 11
questionnaire statements using a 5-point Likert scale (1 = Strongly
disagree, 5 = Strongly agree). The response distribution of the pre- and
post-test were plotted using a Likert plot (Bryer \& Speerschneider,
2016). The statistical significance of the difference between the
post-test and pre-test responses was evaluated through a paired samples
t-test, using Cohen's d as a measure of effect size (d = 0.2 is small, d
= 0.5 is medium, d \textgreater{} 0.8 is large).

The third section contained three open-ended questions that aimed to
capture children\textquotesingle s evolving explanations of the target
concepts through open-ended questions that deliberately asked children
to apply their conceptualizations of social media mechanisms in
real-life situations. These reasoning tasks focused on
children\textquotesingle s awareness and understanding of the variety of
data traces and the mechanism related to profiling, recommending and
engagement, and asked children to explain, in their own words: 1)
\emph{Jarmo uses Instagram daily. He frequently posts various types of
content and actively follows different channels and people. What means
does Instagram use to build an accurate understanding of what Jarmo is
like as a person?} 2) \emph{How do the activities of other users
influence what kinds of content does Instagram recommend to Jarmo?} 3)
\emph{By what means does Instagram get Jarmo to spend more and more time
on it?}.

Approximately one week after the second workshop, students completed
a post-test that did not have the first section but otherwise covered
the two other sections as in the pre-test. In total, 183 students
responded to the pre-test and 191 students responded to the
post-test. While the length of students' answers to open-ended questions
varied, it was usually around one or two sentences.

To analyze the nature of children\textquotesingle s explanations of
the basic mechanisms of social media, the three open-ended
questions in pre/post test data were analyzed using qualitative content
analysis with an inductive approach (Elo \& Kyngäs, 2008). All coding
categories were developed based on the data in collaboration between two
authors, after which the entire data set was coded by one author. To
assess the inter-rater reliability of coding, a third researcher
independently coded 20.2\% of open-ended questions in the data set.
Cohen's kappa statistic was calculated and the overall consensus between
the raters in the first round of coding was found to be 83.2\% (p
\textless{} 0.001), which Landis \& Kock (1977) consider an ``almost
perfect'' agreement. Minor disagreements were discussed and resolved in
consensus between the three researchers. Resulting categories, their
description and example responses are provided in the following section
for readers to consider and evaluate these interpretations (Hammer and
Berland, 2014).

The proportion of responses corresponding to each code was represented
using contiguous stacked bar charts (one for the pre-test and one for
the post-test). The transitions in responses from the pre-test to the
post-test were represented using an alluvial plot between the stacked
bars, created using the R package \emph{ggalluvial} (Brunson, 2020). To
evaluate the association between pre-test and post-test responses we
conducted a chi-square test for each open ended question. We used the
resulting Pearson's residuals to assess the statistical significance
between each transition. Pearson residuals quantify the deviation of
observed frequencies from expected frequencies under the assumption of
independence. We plotted the contingency table between pre- and
post-test using a mosaic plot, where each cell represents the frequency
of participants with each specific combination of pre-test and post-test
responses, and is shaded according to the magnitude and direction of the
Pearson residuals: a blue shade indicates positive residuals and a red
shade indicates negative residuals.

\section{Results}
\subsection{Nature and development of explanations}
\subsubsection{Data traces}

The first open-ended question assessed children's recognition of the
various types of data traces that social media platforms collect from
their users: ''\emph{Jarmo uses Instagram daily. He frequently posts
various types of content and actively follows different channels and
people. What means does Instagram use to build an accurate understanding
of what Jarmo is like as a person?}'' A qualitative content analysis of
all responses in pre- and post-test questions revealed three categories
of recognizing the variety of data traces associated with online
activities (see Appendix). The lowest category (0) included responses
where children either did not answer the question or did not describe
any data traces. These responses accounted for 15.3\% in the pre-test
and 16.2\% in the post-test.

Responses that identified a few data traces but lacked variety were
placed in category 1. This category had the highest frequency of
responses in both the pre-test (74.9\%) and the post-test (52.9\%).
Responses in this category were typically very general, such as
"\emph{it collects pretty much everything of him}'', or they identified
some specific traces, for example, "\emph{so that what kinds of content
Jarmo watches and what content Jarmo posts}''. Many responses mentioned
things like search history, hashtags, viewing history, posts, whom they
are following, and preferred video types. For example, a fifth grader
wrote: "\emph{When Jarmo watches for instance cat videos, he can be
recommended cat videos}"

Compared to categories 0 and 1, responses in category 2 described data
traces and data collection mechanisms with significantly more depth and
detail. For example, in the post-test, one fifth-grader identified
various data traces and their role in recommendations: ``\emph{What
content he watches, what his friends watch, and what Jarmo's interests
are. Also what posts he has put attention, for example, reacted, liked,
followed, commented, or shared. Then Instagram can recommend similar
content to him. And it's easier for Instagram to advertise Jarmo
products he needs or might need. Also if he eagerly follows someone then
Instagram recommends ever more really similar content.}'' While the
pre-test showed scant signs of recognizing the depth and variety of data
traces (9.8\%), the number of responses that explained the traces and
data collection mechanisms more than tripled in the post-test (30.9\%)
(Fig. 7). Some responses in this category also noted the ubiquity of
data collection by referring to data exchange between platforms. For
instance, an eighth-grader wrote: ``\emph{Instagram collects data about
how Jarmo uses the app, which is how long he for example looks at
certain kinds of pictures, does he like them, comment on them, share
others' posts with friends, and so on. Every click and second gives more
data about Jarmo, and based on all the factors one can build an accurate
idea of Jarmo. Also what Jarmo does in other apps and social media
affects, that is, what he Googles, what he buys on web stores, and
everything possible.}''

\begin{figure}
    \centering
\includegraphics[width=\linewidth]{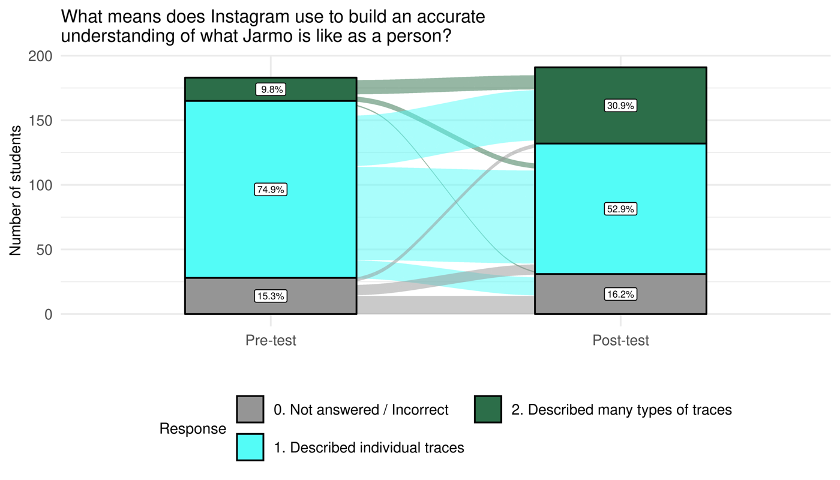}
\caption{Alluvial plot of changes in students' responses between pre-test and post-test to the first open-ended question, which assessed children's recognition of the variety of data traces.}
\label{fig:fig7}
\end{figure}

Figure \ref{fig:fig7} visualizes the transitions in students\textquotesingle{}
responses between the pre-test and post-test. The left bar shows the
pre-test distribution of responses across the categories described above
(see the Appendix for more detailed description of categories and
examples of each). The right bar shows the post-test distribution. The
flows between bars illustrate changes in each child's responses from
pre- to post-test. For example, the topmost light pink flow from
category 1 on the left to category 2 on the right indicates that 39
children whose response was coded as 1 in the pre-test had a response
coded as 2 in the post-test. Pearson residuals show no distinct pattern
of direction of moving between the categories, but the learners' answers
were likely to remain in the same categories (see Figure \ref{fig:figs2} in the
Appendix).

\subsubsection{Profiling and recommending}

The second open-ended question looked into children's awareness and
understanding of social profiling and recommending mechanisms on social
media: ``\emph{How do the activities of other users influence what kinds
of content does Instagram recommend to Jarmo?}'' A qualitative content
analysis of pre- and post-test responses yielded five categories (see
Appendix). The lowest category (0) contained responses where
children either did not answer the question or provided irrelevant
(e.g., not about other users) or clearly incorrect answers. This
category had the highest frequency of responses in both the pre-test
(57.38\%) and post-test (35.60\%). Many responses in this category were
empty or variants of ``\emph{I don\textquotesingle t know}`` answers,
indicating that many children struggled to understand how profiling and
recommending work. In the pre-test many intuitive but incorrect folk
theories were categorized to this category: For instance, one
fifth-grader wrote, ``\emph{Users try to make more content of the kind
Jarmo likes}'' and one eighth-grader wrote, ``\emph{If someone posts a
video that Instagram believes to be interesting to Jarmo, then Instagram
shows it}''. Some post-test answers provided explanations that did not
refer to the role of other users: "\emph{When Jarmo watches Instagram,
Instagram remembers that he has watched car pictures, so it recommends
them to Jarmo}" (5th-grader/post-test).

Responses that recognized the influence of other users at a general
level, but did not draw a connection between other users and Jarmo or
the similarity between the profiles of others and Jarmo's, formed
category 1. The percentage of responses in this category was 22.95\% in
the pre-test and 28.27\% in the post-test. Typically, responses in this
category were very general and touched on aspects like content
similarity ``\emph{If other users' content is similar, then Instagram
recommends them the same things}'' (5th grader/post-test), general
popularity ``\emph{that if something is bigly popular it shows that kind
of content}'' (5th grader/post-test), and interactions such as likes.

Responses that considered the role of data given, such as demographic
data that users typically enter when registering on a platform, were
placed in category 2. These responses typically referred to the age of
the users and without otherwise connecting Jarmo with those other users.
For instance, one 8th grader reasoned in the post-test: ``\emph{If other
people of Jarmo\textquotesingle s age have liked some picture a lot,
it's shown to Jarmo too, because Instagram thinks that Jarmo could like
it too.}'' This category had the lowest frequency of responses in both
the pre-test (0.55\%) and the post-test (2.09\%).

Category 3 consisted of responses that explicitly mentioned interactions
with other users, such as following others and having followers, or
interacting through likes and comments. Compared to the lower
categories, responses in this category showed a more nuanced
understanding of the mechanisms behind recommendations. This category
had 16.39\% of responses in the pre-test, and the share remained
consistent in the post-test, where it had 17.80\% of responses.
Typically, these responses focused on how social activities on the
platform (e.g., tagging, following, links sent by others) influence
recommendations without explicitly referring to profiling or engagement:
''\emph{If someone has created content that Jarmo likes, and if Jarmo
likes or comments or follows, {[}that someone{]} may make more content
like that}'' (5th grader/post-test). Other responses in this category
acknowledged how the actions of others can influence recommendations
Jarmo sees: ``\emph{Jarmo may be recommended videos his, for instance,
followers or whom he follows, watched, commented, or liked}'' (5th
grader/post-test).

Responses in category 4 showed significant detail and depth of
understanding in describing how recommendations are influenced by the
actions of others. These responses considered the similarity of profiles
or users' engagement histories, as one 5th grader explained in the
post-test: ``\emph{The activities of other users similar to Jarmo affect
Instagram recommendations, which Instagram gives Jarmo {[}...{]} what
they do, watch, follow, and so on in Instagram.}'' Some students were
able to explain, in their own words, collaborative filtering and the
association between profiling and recommendations, such as the following
description by an 8th grader in the post-test: ``\emph{For example, if
Jarmo likes cats, comments cat posts, and so on, Instagram starts to
offer Jarmo also for example dog-related content {[}...{]} Because it
infers that Jarmo likes also dogs because he likes cats, and because
other people who like cats have also liked dogs. So when people who like
a similar thing, such as ice hockey which is a sport, they often also
like things of the same type, such as soccer, which is also a sport. And
if Jarmo now follows ice hockey, he also gets offered soccer related
content.}'' Category 4 had few responses in pre-test (2.73\%), but the
percentage of students who provided advanced and more scientifically
accurate explanations of profiling and recommendation mechanisms
increased sixfold in the post test (16.23\%).

\begin{figure}
    \centering
    \includegraphics[width=\linewidth]{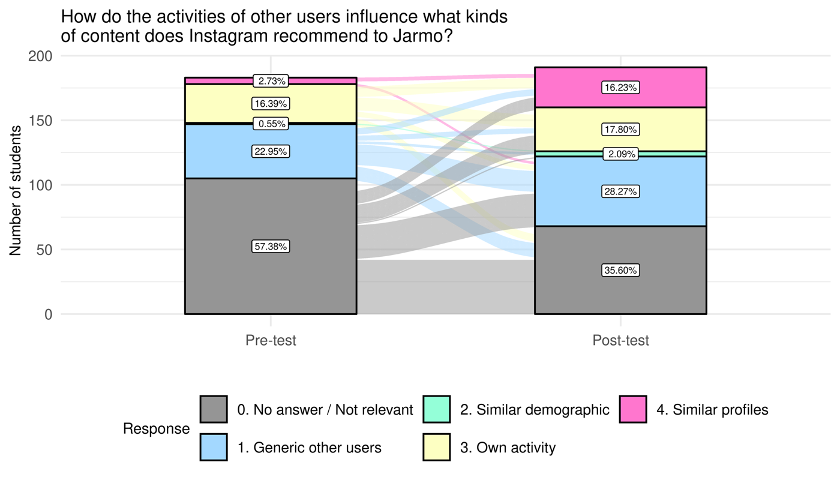}
    \caption{Alluvial plot of changes in students' responses between pre-test and post-test to the second open-ended question, which assessed
children's awareness and understanding of social profiling and
recommending mechanisms on social media.}
    \label{fig:fig8}
\end{figure}

Figure \ref{fig:fig8} visualizes the transitions in students\textquotesingle{}
responses between the pre- and post-test on profiling and recommending
mechanisms. A notable shift can be seen away from Category 0, with the
majority of changes moving to Category 1 (generic ``other users''
without making a connection to their similarity with
Jarmo\textquotesingle s own activities or profile). There is also a
notable increase in responses that refer to the similarity of profiles,
activities, or engagement patterns between Jarmo and other users,
demonstrating an understanding of co-engagement, profile similarity, or
other key socially driven mechanisms of recommending. Pearson residuals
show that students who were in categories 2, 3 and 4 in the pre-test
were more likely to remain therein in the post-test, while students in
categories 0 and 1 showed no statistically significant distinct pattern
of direction of moving between the categories (see Figure \ref{fig:figs4} in the
Appendix).

\subsubsection{Engagement}

The third open-ended question probed children's understanding of
engagement related mechanisms on social media: ``\emph{By what means
does Instagram get Jarmo to spend more and more time on it?}'' A
qualitative content analysis of pre/post data yielded five categories.
The first category (0) consisted of responses where children either did
not answer the question or provided irrelevant answers with no
explanations, such as ''\emph{when he gets addicted}'' (5th
grader/pre-test). This category accounted for 22.40\% of the pre-test
responses, with a notable drop to 13.09\% in the post-test.

Responses in category 1 included generic descriptions related to
recommending, such as "good," "addictive," "showing more content'' or
``sending announcements.'' Typically, these answers were made at a
general level, like \emph{``{[}Instagram{]} puts addictive videos and
advertisement}'' (5th grader/post-test). While this category indicated a
rudimentary awareness of some building blocks of targeting or
engagement, responses in category 1 lacked specific explanations or
references to their underlying mechanisms. The share of category 1
responses remained relatively consistent, accounting for 26.78\% in the
pre-test and 25.13\% in the post-test.

Responses in category 2 included some references to targeting, such as
content that the user likes, prefers, or is interested in, but without
explicit reference to mechanisms of data collection or engagement. Many
of these responses mentioned the concept of recommending, like one 5th
grader reasoned in the post-test: ``\emph{by recommending Jarmo those
kinds of pictures that Jarmo is interested in}.'' This category had the
highest frequency of responses in both the pre-test (40.98\%) and the
post-test (46.07\%), indicating that to some extent, the children
understood that systems can selectively target content, but had not
connected this to the deeper, data-driven mechanism of tracking,
profiling, or engagement.

Responses in category 3 also referenced targeting, but they differed
from category 2 responses by explaining how one's data traces or
inferred data are used to target content and enhance engagement. These
responses showed a more nuanced understanding, whereby children not only
wrote that the content recommendations are tailored to user preferences
but also that this personalization is data-driven. For instance,
children mentioned how recommendations are based on viewing history,
such as ``\emph{Tracking what Jarmo watches and recommending more good
videos.}'' (5th-grader/post-test) or like and comment history:
``\emph{Recommending similar content that he has liked and commented on
already earlier}'' (8th-grader/post-test). These responses, indicating a
more advanced comprehension of how personal data are used for increasing
engagement and targeting content, accounted for 9.29\% of the responses
in the pre-test and 6.28\% of the responses in the post-test. This
suggests that many children were already aware of some of the mechanics
behind content recommendation systems, understanding how one's content
feed is personalized but also how users' actions, data given, and data
traces on the platform contribute to that personalization.

Responses in category 4 demonstrated an advanced understanding of
engagement mechanisms by referring to targeting based on emotional
reactions, provocations, randomness, (co-)engagement, and profiling. For
instance, one fifth-grader explained in the post-test how social media
platforms might increase Jarmo's engagement by suggesting new things, by
recommending content that Jarmo consistently likes, and showing content
popular among users who are similar to Jarmo: ``\emph{Instagram can make
Jarmo use Instagram more, for example, by suggesting Jarmo new things,
recommending him things he has always liked, showing other things that
users like him have liked, and so on}`` (5th-grader/post-test). Some
responses captured how content not only aligns with user preferences but
also how it is engineered to elicit strong emotional responses, further
driving engagement. For example, one eighth-grader articulated this in
the post-test by reasoning as follows: ``\emph{Instagram aims to give
Jarmo content that he likes, so that this way Jarmo would {[}click{]}
like them and comment and watch Instagram more. Instagram can also post
Jarmo content that he hates so he would then comment on them and that
way use Instagram more. Instagram is also designed to be as addictive as
possible so that Jarmo wouldn't want to stop watching it''}. Responses
in category 4 were nearly absent in the pre-test (0.55\%), but the
frequency of such advanced reasoning increased in the post-test to
9.42\%.

\begin{figure}
\centering
\includegraphics[width=\linewidth]{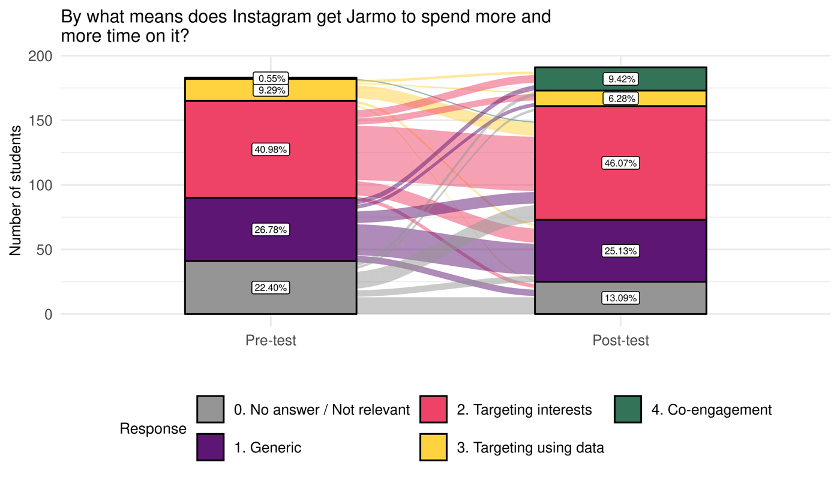}
\caption{Alluvial plot of changes in students' responses between pre-test and post-test to the third open-ended question, which assessed children's awareness and understanding of engagement related mechanisms
on social media.}
\label{fig:fig9}
\end{figure}

Figure \ref{fig:fig9} shows the transitions in responses to the question related to
engagement mechanisms on social media. Many students whose answers had
been in categories 0, 1, and 2 shifted to categories 2, 3, and 4.
Categories 3 and 4, which showed data-driven or engagement-related
explanations, grew slightly. However, the question on engagement showed
clearly fewer advanced, data-driven explanations compared to the other
two open-ended questions. Pearson residuals show that students in
categories 0, 1 and 2 on the pre-test were more likely to remain therein
on the post-test, whereas for students in categories 3 and 4 no
transition was statistically significant (see Fig. S\ref{fig:figs6} on the Appendix).

\subsection{Perceptions of one's own data agency}
The results on the 11-point data agency self-evaluation (Authors,
forthcoming) instrument showed positive impact across all the
statements, with medium effect size (d=0.63***) on statements on
understanding how social media collects information about users (the
number of positive responses increased from 65\% to 91\%) and on
recognition that social media platforms aim to maximize their use (the
number of positive responses increased from 65\% to 87\%, d=0.54***) (see Fig. \label{fig:likert}).
\begin{figure}
    \centering
    \includegraphics[width=.99\linewidth]{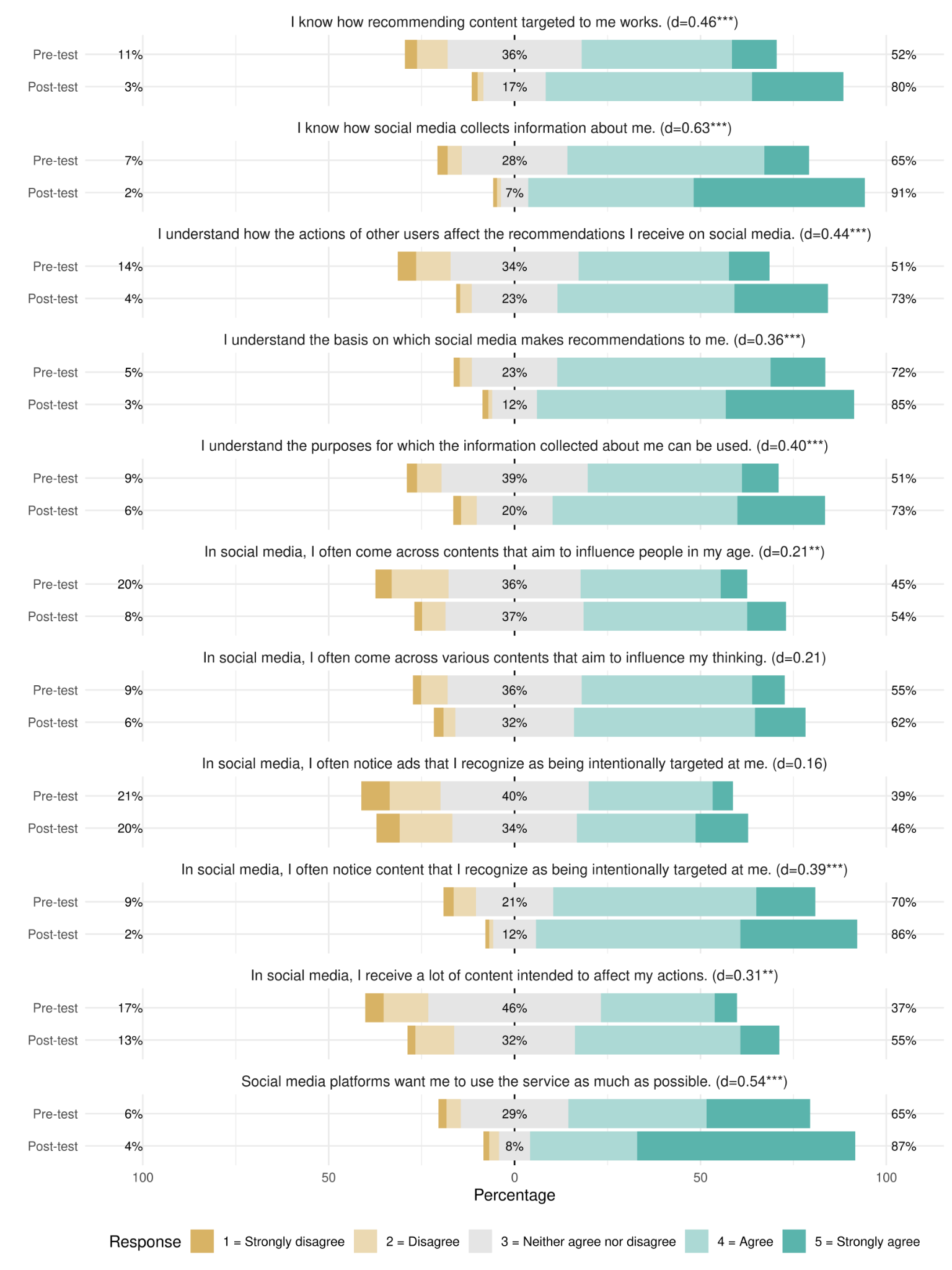}
    \caption{Impact of intervention on participants' perceptions of
their data agency (pre- / post-test) (Authors, forthcoming).}
    \label{fig:likert}
\end{figure}

A medium effect size was observed across several statements: For
example, perception of knowing how content recommendation works
increased (the number of positive responses increased from 52\% to 80\%,
d=0.46***), as did sense of awareness of the impact of other users'
actions on personal recommendations (the number of positive responses
increased from 51\% to 73\%, d=0.44***) and sense of knowledge of the
purposes for which collected information can be used (the number of
positive responses increased from 51\% to 73\%, d=0.40***). The rest
ranged from small to medium effect, with statistically significant
improvement in all but two statements (see Fig.  \ref{fig:likert}).

\section{Discussion}

To strengthen young learners' data agency and digital literacy, this
study set out to investigate how to facilitate their awareness and
understanding of the basic concepts and mechanisms of social media and
online platforms. This article describes pedagogical insights, learning
activities, and educational technology designed to foster collaborative
learning and critical inquiry that afforded children to engage in
hands-on exploration of the basic data-driven concepts and principles
central to how social media works. The intervention aimed to make
concrete a number of abstract concepts in the domain of data-driven
computing.

To answer the first research question, ``\emph{How do children's
explanations of the basic mechanisms of social media develop over a
four-hour technology-rich educational intervention?}'', the intervention
presented students with concrete examples of three types of data and how
they are generated---data given (volunteered data), data traces
(observed data), and inferred data (van der Hof, 2016; Pangrazio \&
Selwyn, 2019). The results show a quantitative and qualitative
transition from a limited number of general descriptions of data traces
to a larger number of more varied and more detailed descriptions. This
meets the study's aim to improve students' understanding of how data are
collected and used to generate insights about users.

Moving from data to profiling, the results show more than twenty percent
drop in incorrect answers (57.38\% to 35.60\%) and a clear qualitative
improvement in students' understanding of social-based profiling and
recommending mechanisms. The results suggest an ability to bridge the
gap from everyday experiences to science-based understanding. These
results are consistent with those of Vosniadou et al. (2008), which
showed that collaborative activities, joint exploration, dialogical
interaction, and classroom discussions can help students move from their
initial misconceptions to disciplinary ways of thinking. While the
results on engagement mechanisms were less clear, the number of answers
showing advanced understanding grew from nearly zero to 9.42\%, and the
number of incorrect answers dropped by nine percent (22.40\% to
13.09\%).

Although the intervention was short, only two sessions, the results show
that hands-on activities with real-life examples can facilitate
conceptual change from students' initial ideas to more sophisticated and
scientific explanations, even with a complex phenomenon like social
media mechanisms (Aleknavičiūtė et al., 2023; Lehtinen et al., 2020).
However, the development of deeper level understandings of scientific
concepts is a long and gradual process, because it requires significant
re-organization of existing knowledge structures (Vosniadou, 2013).
Thus, this highlights the importance of educational practices that
engage students in sustained advancement of understanding, for example,
by revisiting the same social media concepts and their complex
relationships in varying contexts, from a range of perspectives and
through new and deeper inquiry questions as old ones are adressed.

Regarding the second research question, ``\emph{How do children's
perceptions of their data agency in relation to social media develop
over that intervention?}'', the results showed a statistically
significant positive impact across the statements, with varying effect
sizes. Most importantly, there was significant improvement on statements
concerning understanding how social media collects information about
users and recognizing that social media platforms aim to maximize their
use (engagement). This underscores the importance of designing
educational practices and technologies that not only support the
development of conceptual understanding but also cultivates
students\textquotesingle{} curiosity, interest, and sense of agency in
learning ---as suggested by educational research (e.g. Blumenfeld et
al., 1991; Krajcik, \& Blumenfeld, 2006; Hakkarainen \&
Seitamaa-Hakkarainen, 2022). By doing so, students are also scaffolded
to develop volition and motivation to revisit and reconstruct their
initial conceptions and to build a more deeper and critical
understanding of the social media mechanisms that they interact with in
everyday life.

These results contribute both theoretical and empirical insights for
supporting the learning of computing's disciplinary concepts and
mechanisms related to the functioning of social media. Theoretically,
the study poses a complex question of how to promote scientific
understanding of mechanisms that are epistemically opaque by nature.
These new relationships and dynamics highlight the need for frameworks
in conceptual change that take into account the new kind of epistemic
opacity that mediates human agency, learning, and the formation of
everyday folk theories about how the world works.

While there is much literature theorizing and showcasing the
difficulties in supporting students' conceptual change (e.g. Chi, 2008;
Vosniadou et al., 2008, Aleknavičiūtė et al., 2023), this study
contributes to this empirical body of research by illustrating the
nature and levels of children\textquotesingle s conceptions and the
ontological shifts needed to progress towards a more disciplinary
understanding of social media. This includes moving from viewing social
media as simple communication platforms to recognizing them as
sophisticated systems designed to collect data, profile users, and
tailor content dynamically to engage users.

Furthermore, the results showed that children have formulated their own
folk theories of computational systems through everyday use of social
media (see DeVito et al., 2018), which can create barriers to adopting
more scientifically accurate conceptions. This is particularly so in
complex and opaque areas like data collection, profiling, and
recommendation algorithms. Understanding the nature and levels of
children\textquotesingle s conceptions offers important pedagogical
implications and insights for scaffolding children\textquotesingle s
learning and understanding. By becoming more aware of these conceptual
levels and challenges, teachers can more effectively tailor their
pedagogical strategies to support students in reconstructing their naïve
conceptions and build a more accurate and scientific understanding of
social media mechanisms.

In terms of scaffolding, the results also confirm that computer
simulations and games, combined with collaborative, inquiry-based
learning, can support children in developing a more scientific
understanding of the world (Aleknavičiūtė et al., 2023; Osterhaus et
al., 2021, Smetana \& Bell, 2012; Trundle \& Bell, 2010). Scaffolding
provided through multiple affordances---such as curriculum materials,
tools, educational technology, peer discussions, and teacher
facilitation---is essential not only for conceptual change but also for
cultivating children\textquotesingle s agentive actions and abilities to
make informed decisions and actions in the algorithmic culture they
inhabit.

As with any study, this research is limited in several ways.
Firstly, the interventions were implemented in a particular cultural
setting, influenced by context-bound social interactions and specific
tool-mediated actions. Secondly, the data were collected from a
specific cohort of school students in Finland. Thirdly, the mechanisms
presented by Somekone, while not incorrect, are a simplification of
reality, whereas most platforms use more complicated variants of machine
learning.

To address the first two limitations and enhance the
generalizability of the findings, future research should implement
interventions and collect data from a broader range of children and
young people, encompassing diverse cultural backgrounds, educational
settings, and children with a variety of educational needs. Future
research should also explore the longer-term impact of sustained
interventions that present social media's mechanisms from different
angles and in different contexts. They could expand the intervention to
older age groups, such as high school students, to study what new
computational concepts can be included at higher levels of education
(e.g., Rizvi et al., 2023). In order to enable building better
professional development activities for teachers (e.g., Olari et al.,
2024), there is also an evident need to study the specific disciplinary
knowledge, pedagogical competencies, and technological skills they need
for scaffolding students\textquotesingle{} learning (e.g., Grover, 2024;
Druga et al., 2022; Long \& Magerko, 2020).

\section{REFERENCES}
\bibliographystyle{ACM-Reference-Format}

Acquisti, A., Brandimarte, L., and Loewenstein, G. (2015). Privacy and
human behavior in the age of information. Science, 347(6221):509--514.

Aleknavičiūtė, V., Lehtinen, E., \& Södervik, I. (2023). Thirty years of
conceptual change research in biology -- A review and meta-analysis of
intervention studies. \emph{Educational Research Review}, \emph{41},
100556. \url{https://doi.org/10.1016/j.edurev.2023.100556}

Barassi, V. (2020). Child \textbar{} data \textbar{} citizen: How tech
companies are profiling us from before birth. The MIT Press, Cambridge,
MA, USA.

Benjamin, R. (2019). Race After Technology: Abolitionist Tools for the
New Jim Code. Polity, Cambridge, MA, USA.

Blumenfeld, P. C., Soloway, E., Marx, R. W., Krajcik, J. S., Guzdial,
M., \& Palincsar, A. (1991). Motivating Project-Based Learning:
Sustaining the Doing, Supporting the Learning. Educational Psychologist,
26(3--4), 369--398.
\url{https://doi.org/10.1080/00461520.1991.9653139}

Bowker, G. C. and Star, S. L. (2000). Sorting Things Out: Classification
and Its Consequences. The MIT Press, Cambridge, MA, USA.

Bowler, L., Acker, A., Jeng, W., and Chi, Y. (2017). ``It lives all
around us'': Aspects of data literacy in teen's lives. Proceedings of
the Association for Information Science and Technology, 54(1):27--35.

boyd, d. m. and Ellison, N. B. (2007). Social network sites: Definition,
history, and scholarship. Journal of Computer-Mediated Communication,
13(1):210--230.

boyd, d. m. (2015). It's Complicated: The Social Lives of Networked
Teens. Yale University Press, New Haven, CT, USA.

Brunson, J. (2020). ggalluvial: Layered Grammar for Alluvial Plots.
\emph{Journal of Open Source Software, 5}(49).
\url{https://doi.org/10.21105/joss.02017}

Bryer, J., \& Speerschneider, K. I. (2016). \emph{likert: Analysis and
Visualization Likert Items} (Version 1.3.5) {[}R{]}.
\url{https://CRAN.R-project.org/package=likert}

Chi, M. (2008). Three types of conceptual change: Belief revision,
mental model transformation, and categorical shift. In S. Vosniadou
(Ed.), \emph{Handbook of research on conceptual change} (pp. 61--82).
Erlbaum.

Chiu, T. K., Ahmad, Z., Ismailov, M., and Sanusi, I. T. (2024). What are
artificial intelligence literacy and competency? A comprehensive
framework to support them. Computers and Education Open, 6:100171.

Crawford, K. (2021). Atlas of AI: Power, Politics, and the Planetary
Costs of Artificial Intelligence. Yale University Press, New Haven, CT,
USA.

DeVito, M. A., Birnholtz, J., Hancock, J. T., French, M., \& Liu, S.
(2018). How People Form Folk Theories of Social Media Feeds and What it
Means for How We Study Self-Presentation. \emph{Proceedings of the 2018
CHI Conference on Human Factors in Computing Systems}, 1--12.
\url{https://doi.org/10.1145/3173574.3173694}

DeVito, M. A., Gergle, D., \& Birnholtz, J. (2017). ``Algorithms ruin
everything'': \#RIPTwitter, Folk Theories, and Resistance to Algorithmic
Change in Social Media. \emph{Proceedings of the 2017 CHI Conference on
Human Factors in Computing Systems}, 3163--3174.
\url{https://doi.org/10.1145/3025453.3025659}

Dolan, R., Conduit, J., Fahy, J., and Goodman, S. (2016). Social media
engagement behaviour: a uses and gratifications perspective. Journal of
Strategic Marketing, 24(3-4):261--277.

Druga, S., Otero, N., and Ko, A. J. (2022). The landscape of teaching
resources for AI education. In Proceedings of the 27th ACM Conference on
on Innovation and Technology in Computer Science Education Vol. 1,
ITiCSE '22, pages 96--102, New York, NY, USA. ACM.

Dwivedi, Y. K., Ismagilova, E., Rana, N. P., and Raman, R. (2023).
Social media adoption, usage and impact in business-to-business (B2B)
context: A state-of-the-art literature review. \emph{Information Systems
Frontiers}, 25(3):971--993.

Elo, S., \& Kyngäs, H. (2008). The qualitative content analysis process.
\emph{Journal of Advanced Nursing}, \emph{62}(1), 107--115.
\url{https://doi.org/https://doi.org/10.1111/j.1365-2648.2007.04569.x}

Eubanks, V. (2018). Automating Inequality: How High-Tech Tools Profile,
Police, and Punish the Poor. St. Martin's Press, New York, NY, USA.

Fisher, M. (2022). \emph{The Chaos Machine: The Inside Story of How
Social Media Rewired Our Minds and Our World}. Little, Brown and
Company.

Freire, P. (1974). \emph{Pedagogy of the oppressed}. Seabury Press.

Grover, S. (2024). Teaching AI to K-12 learners: Lessons, issues, and
guidance. In Proceedings of ACM Computer Science Education (SIGCSE) 2024
Conference, pages 1--7, Portland, OR, USA. ACM.\\
\url{https://doi.org/10.1145/3626252.3630937}

Hakkarainen, K., \& Seitamaa-Hakkarainen, P. (2022). Learning by
inventing\,: Theoretical foundations. In T. Korhonen, K. Kangas, \& L.
Salo (Eds.), \emph{Invention pedagogy\,: The Finnish approach to maker
education} (pp. 15--27). Routledge.

Hallinan, B., \& Striphas, T. (2016). Recommended for you: The Netflix
Prize and the production of algorithmic culture. \emph{New Media and
Society}, \emph{18}(1), 117--137.
\url{https://doi.org/10.1177/1461444814538646}

Hammer, D., \& Berland, L. K. (2014). Confusing Claims for Data: A
Critique of Common Practices for Presenting Qualitative Research on
Learning. \emph{Journal of the Learning Sciences}, \emph{23}(1), 37--46.
\url{https://doi.org/10.1080/10508406.2013.802652}

Heintz, F. and Roos, T. (2021). Elements of AI - teaching the basics of
AI to everyone in Sweden. In Proceedings of the 13th International
Conference on Education and New Learning Technologies (EDULEARN21),
pages 2568--2572, Online. IATED.

Hendricks, V. F., \& Vestergaard, M. (2018). Reality lost: Markets of
attention, misinformation and manipulation. In \emph{Reality Lost:
Markets of Attention, Misinformation and Manipulation}.
\url{https://doi.org/10.1007/978-3-030-00813-0}

Höper, L. and Schulte, C. (2024). Empowering students for the
data-driven world: A qualitative study of the relevance of learning
about data-driven technologies. Informatics in Education.
\url{https://doi.org/10.15388/infedu.2024.19}

Jünger, J. (2018). Mapping the field of automated data collection on the
web: Collection approaches, data types, and research logic. In Stuetzer,
C. M., Welker, M., and Egger, M., editors, Computational Social Science
in the Age of Big Data: Concepts, Methodologies, Tools, and
Applications, pages 104--130. Herbert von Halem Verlag, Köln, Germany.

Kafai, Y. B., Fields, D. A., \& Searle, K. A. (2018). Understanding
Media Literacy and DIY Creativity in Youth Digital Productions. In
\emph{The International Encyclopedia of Media Literacy}.
\url{https://doi.org/10.1002/9781118978238.ieml0058}

Karizat, N., Delmonaco, D., Eslami, M., and Andalibi, N. (2021).
Algorithmic folk theories and identity: How TikTok users co-produce
knowledge of identity and engage in algorithmic resistance. Proceedings
of the ACM on Human-Computer Interaction, 5(CSCW2).
\url{https://doi.org/10.1145/3476046}

Kitchin, R. (2012). The Data Revolution: Big Data, Open Data, Data
Infrastructures and Their Consequences. In \emph{Online Information
Review} (Vol. 39, Issue 2). Sage.
\url{https://doi.org/10.1108/oir-01-2015-0011}

Klinger, U., \& Svensson, J. (2018). The end of media logics? On
algorithms and agency. \emph{New Media and Society}, \emph{20}(12),
4653--4670. \url{https://doi.org/10.1177/1461444818779750}

Koskinen, I. (2023). We have no satisfactory social epistemology of
AI-based science. Social Epistemology, pages 1--18.
\url{https://doi.org/10.1080/02691728.2023.2286253}

Krajcik, J. S., \& Blumenfeld, P. C. (2006). Project-based learning. In
\emph{The Cambridge handbook of the learning sciences} (pp. 317--333).
Cambridge University Press.

Kramer, A. D. I., Guillory, J. E., \& Hancock, J. T. (2014).
Experimental evidence of massive-scale emotional contagion through
social networks. \emph{Proceedings of the National Academy of Sciences
of the United States of America}, \emph{111}(24), 8788--8790.
\url{https://doi.org/10.1073/pnas.1320040111}

Landis, J. R., \& Koch, G. G. (1977). The Measurement of Observer
Agreement for Categorical Data. \emph{Biometrics}, \emph{33}(1),
159--174. \url{https://doi.org/10.2307/2529310}

Lehtinen, E., Gegenfurtner, A., Helle, L., \& Säljö, R. (2020).
Conceptual change in the development of visual expertise.
\emph{International Journal of Educational Research}, \emph{100}.
\url{https://doi.org/10.1016/j.ijer.2020.101545}

Livingstone, S., Stoilova, M., \& Nandagiri, R. (2019). \emph{Talking to
children about data and privacy online: research methodology}.
\url{https://www.lse.ac.uk/media-and-communications/assets/documents/research/projects/childrens-privacy-online/Talking-to-children-about-data-and-privacy-online-methodology-final.pdf}

Long, D. and Magerko, B. (2020). What is AI literacy? competencies and
design considerations. In Proceedings of the 2020 CHI Conference on
Human Factors in Computing Systems, CHI '20, pages 1--16, New York, NY,
USA. ACM. \url{https://doi.org/10.1145/3313831.3376727}

Martins, R. M. and Gresse Von Wangenheim, C. (2022). Find- ings on
teaching machine learning in high school: A ten - year systematic
literature review. Informatics in Education.
\url{https://doi.org/10.15388/infedu.2023.18}

McCosker, A. (2017). Data literacies for the postdemographic social
media self. \emph{First Monday}. \url{https://doi.org/10.5210/fm.v22i10.7307}

Mertala, P., Fagerlund, J., \& Calderon, O. (2022). Finnish 5th and 6th
grade students' pre-instructional conceptions of artificial intelligence
(AI) and their implications for AI literacy education. \emph{Computers
and Education: Artificial Intelligence}, \emph{3}, 100095.
\url{https://doi.org/10.1016/j.caeai.2022.100095}

Morales-Navarro, L., Kafai, Y. B., Nguyen, H., DesPortes, K., Vacca, R.,
Matuk, C., Silander, M., Amato, A., Woods, P., Castro, F., Shaw, M.,
Akgun, S., Greenhow, C., and Garcia, A. (2024). Learning about data,
algorithms, and algorithmic justice on TikTok in personally meaningful
ways. In \emph{Proceedings of the 18th International Conference of the
Learning Sciences}, ICLS 2024. International Society of the Learning
Sciences.

Moritz Büchi Eduard Fosch-Villaronga, C. L. A. T.-L., \& Velidi, S.
(2023). Making sense of algorithmic profiling: user perceptions on
Facebook. \emph{Information, Communication \& Society}, \emph{26}(4),
809--825. \url{https://doi.org/10.1080/1369118X.2021.1989011}

Mühling A and Große-Bölting G (2023) Novices' conceptions of machine
learning. Computers and Education: Artificial Intelligence 4: 100142.
\url{https://doi.org/10.1016/j.caeai.2023.100142}

Noble, S. U. (2018). Algorithms of Oppression: How Search Engines
Reinforce Racism. New York University Press, New York, NY, USA.

Ofcom (2023). Children's media lives 2023: A report for Ofcom. Office of
Communications, London, UK.
\url{https://www.ofcom.org.uk/\_\_data/assets/pdf\_file/0025/255850/childrens-media-lives-2023-summary-report.pdf}

Olari, V., Zoppke, T., Reger, M., Samoilova, E., Kandlhofer, M.,
Dagiene, V., Romeike, R., Lieckfeld, A. S., and Lucke, U. (2024).
Introduction of artificial intelligence literacy and data literacy in
computer science teacher education. In Proceedings of the 23rd Koli
Calling International Conference on Computing Education Research, Koli
Calling '23, New York, NY, USA. ACM.
\url{https://doi.org/10.1145/3631802.3631851}

Osterhaus, C., Brandone, A. C., Vosniadou, S., \& Nicolopoulou, A.
(2021). Editorial: The Emergence and Development of Scientific Thinking
During the Early Years: Basic Processes and Supportive Contexts. In
\emph{Frontiers in Psychology} (Vol. 12).
\url{https://doi.org/10.3389/fpsyg.2021.629384}

Pangrazio, L., \& Selwyn, N. (2019). `Personal data literacies': A
critical literacies approach to enhancing understandings of personal
digital data. \emph{New Media and Society}, \emph{21}(2), 419--437.
\url{https://doi.org/10.1177/1461444818799523}

Rasi, P., Vuojärvi, H., \& Ruokamo, H. (2019). Media literacy education
for all ages. \emph{Journal of Media Literacy \ldots{}}.

Rizvi, S., Waite, J., and Sentance, S. (2023). Artificial intelligence
teaching and learn- ing in K-12 from 2019 to 2022: A systematic
literature review. Computers and Education: Artificial Intelligence,
4:100145.

Schlatter, E., Molenaar, I., \& Lazonder, A. W. (2020). Individual
Differences in Children's Development of Scientific Reasoning Through
Inquiry-Based Instruction: Who Needs Additional Guidance?
\emph{Frontiers in Psychology}, \emph{11}.
\url{https://doi.org/10.3389/fpsyg.2020.00904}

Shapiro, R. B. and Fiebrink, R. (2019). Introduction to the special
section: Launching an agenda for research on learning machine learning.
ACM Transactions on Computing Education, 19(4):30:1--30:6.

Smetana, L. K., \& Bell, R. L. (2012). Computer Simulations to Support
Science Instruction and Learning: A critical review of the literature.
\emph{International Journal of Science Education}, \emph{34}(9),
1337--1370. \url{https://doi.org/10.1080/09500693.2011.605182}

Smith, H. J., Dinev, T., and Xu, H. (2011). Information privacy
research: An interdisciplinary review. MIS Quarterly, 35(4):989--1015.

Stoilova, M., Livingstone, S., \& Nandagiri, R. (2020). Digital by
default: Children's capacity to understand and manage online data and
privacy. \emph{Media and Communication}, \emph{8}(4), 197--207.
\url{https://doi.org/10.17645/mac.v8i4.3407}

Swan, M. (2012). Sensor mania! the Internet of Things, wearable
computing, objective metrics, and the quantified self 2.0. Journal of
Sensor and Actuator networks, 1(3):217--253.

TENK. (2019). \emph{The ethical principles of research with human
participants and ethical review in the human sciences in Finland Finnish
National Board on Research Integrity TENK guidelines 2019.}
\url{https://tenk.fi/sites/default/files/2021-01/Ethical\_review\_in\_human\_sciences\_2020.pdf}

Trundle, K. C., \& Bell, R. L. (2010). The use of a computer simulation
to promote conceptual change: A quasi-experimental study.
\emph{Computers \& Education}, \emph{54}(4), 1078--1088.
\url{https://doi.org/10.1016/j.compedu.2009.10.012}

Valkenburg, P. M., Meier, A., and Beyens, I. (2022). Social media use
and its impact on adolescent mental health: An umbrella review of the
evidence. Current Opinion in Psychology, 44:58--68.

Valtonen, T., Tedre, M., Mäkitalo, K., \& Vartiainen, H. (2019). Media
Literacy Education in the Age of Machine Learning. \emph{Journal of
Media Literacy Education}, \emph{11}(2).
\url{https://doi.org/10.23860/jmle-2019-11-2-2}

van der Hof, S. (2017). I agree, or do I? A rights-based analysis of the
law on children's consent in the digital world. Wisconsin International
Law Journal, 34(2):409--445.

van Dijck, J. (2013). ``You have one identity'': Performing the self on
Facebook and LinkedIn. \emph{Media, Culture and Society}, \emph{35}(2),
199--215. \url{https://doi.org/10.1177/0163443712468605}

Van Mechelen, M., Smith, R. C., Schaper, M.-M., Tamashiro, M., Bilstrup,
K.-E., Lunding, M., Graves Petersen, M., and Sejer Iversen, O. (2023).
Emerging technologies in K--12 education: A future HCI research agenda.
\emph{ACM Transactions on Computer-Human Interaction}, 30(3).
\url{https://doi.org/10.1145/3569897}

Vartiainen, H. and Tedre, M. (2024). How text-to-image generative AI is
transforming mediated action. IEEE Computer Graphics and Applications,
pages 1--12. \url{https://doi.org/10.1109/MCG.2024.3355808}

Vartiainen, H., Kahila, J., Tedre, M., López-Pernas, S., and Pope, N.
(2024). Enhancing children's understanding of algorithmic biases in and
with text-to-image generative AI. New Media \& Society.
\url{https://doi.org/10.1177/14614448241252820}

Vosniadou, S. (2007). The cognitive-situative divide and the problem of
conceptual change. \emph{Educational Psychologist}, \emph{42}(1).
\url{https://doi.org/10.1080/00461520709336918}

Vosniadou, S. (2013). Conceptual change in learning and instruction: The
framework theory approach. In \emph{International Handbook of Research
on Conceptual Change}. \url{https://doi.org/10.4324/9780203154472}

Vosniadou, S., Vamvakoussi, X., \& Skopeliti, I. (2008). The framework
theory approach to the problem of conceptual change. In
\emph{International handbook of research on conceptual change} (Issue
February 2019).

Zuboff, S. (1988). In the Age of the Smart Machine: The Future of Work
and Power. Basic Books, New York, NY, USA.

Zuboff, S. (2015). Big other: Surveillance capitalism and the prospects
of an information civilization. \emph{Journal of Information
Technology}, \emph{30}(1), 75--89. \url{https://doi.org/10.1057/jit.2015.5}

Zuboff, S. (2019). The Age of Surveillance Capitalism: The Fight for a
Human Future at the New Frontier of Power. Profile Books, New York, NY,
USA.

\appendix

\renewcommand{\thetable}{S\arabic{table}}
\renewcommand{\thefigure}{S\arabic{figure}}

\setcounter{table}{0}
\setcounter{figure}{0}
\begin{table}[ht]
  \centering
    \caption{Number and proportion of students of each response per test and grade for the first open-ended question}

  \begin{tabular}{@{}
    >{\raggedright\arraybackslash}p{0.25\textwidth} 
    >{\raggedright\arraybackslash}p{0.10\textwidth} 
    >{\raggedright\arraybackslash}p{0.10\textwidth} 
    >{\raggedright\arraybackslash}p{0.10\textwidth} 
    >{\raggedright\arraybackslash}p{0.10\textwidth} 
    >{\raggedright\arraybackslash}p{0.10\textwidth} 
    >{\raggedright\arraybackslash}p{0.10\textwidth} @{}}
    \toprule
    & \textbf{Pre-test Grade 5} & \textbf{Pre-test Grade 8} & \textbf{Post-test Grade 5} & \textbf{Post-test Grade 8} & \textbf{Pre-test Total} & \textbf{Post-test Total} \\
    \midrule
    \textbf{0: No answer, "I don't know", or answers that do not address the question.} & \makecell{17 \\ (14.53\%)} & \makecell{11 \\ (16.67\%)} & \makecell{23 \\ (19.01\%)} & \makecell{8 \\ (11.43\%)} & \makecell{28 \\ (15.3\%)} & \makecell{31 \\ (16.23\%)} \\
    \textbf{1: Answer that identifies generic "data" or names one or two data traces.} & \makecell{93 \\ (79.49\%)} & \makecell{44 \\ (66.67\%)} & \makecell{66 \\ (54.55\%)} & \makecell{35 \\ (50\%)} & \makecell{137 \\ (74.86\%)} & \makecell{101 \\ (52.88\%)} \\
    \textbf{2: Answer that names many data traces.} & \makecell{7 \\ (5.98\%)} & \makecell{11 \\ (16.67\%)} & \makecell{32 \\ (26.45\%)} & \makecell{27 \\ (38.57\%)} & \makecell{18 \\ (9.84\%)} & \makecell{59 \\ (30.89\%)} \\
    \bottomrule
  \end{tabular}
  \label{s1}
\end{table}

\begin{table}[ht]
  \centering
    \caption{Number and proportion of students of each response per test and grade for the second open-ended question}

  \begin{tabular}{@{}
    >{\raggedright\arraybackslash}p{0.25\textwidth} 
    >{\raggedright\arraybackslash}p{0.10\textwidth} 
    >{\raggedright\arraybackslash}p{0.10\textwidth} 
    >{\raggedright\arraybackslash}p{0.10\textwidth} 
    >{\raggedright\arraybackslash}p{0.10\textwidth} 
    >{\raggedright\arraybackslash}p{0.10\textwidth} 
    >{\raggedright\arraybackslash}p{0.10\textwidth} @{}}
    \toprule
    & \textbf{Pre-test Grade 5} & \textbf{Pre-test Grade 8} & \textbf{Post-test Grade 5} & \textbf{Post-test Grade 8} & \textbf{Pre-test Total} & \textbf{Post-test Total} \\
    \midrule
    \textbf{0: No answer, "I don't know", or answers that do not address the question.} & \makecell{28 \\ (23.93\%)} & \makecell{13 \\ (19.7\%)} & \makecell{16 \\ (13.22\%)} & \makecell{9 \\ (12.86\%)} & \makecell{41 \\ (22.4\%)} & \makecell{25 \\ (13.09\%)} \\
    \textbf{1: The answer is limited to other users' activities without connecting them to their similarity with Jarmo's own activities or profile.} & \makecell{36 \\ (30.77\%)} & \makecell{13 \\ (19.7\%)} & \makecell{35 \\ (28.93\%)} & \makecell{13 \\ (18.57\%)} & \makecell{49 \\ (26.78\%)} & \makecell{48 \\ (25.13\%)} \\
    \textbf{2: The answer refers to similar demographics between the others and Jarmo.} & \makecell{46 \\ (39.32\%)} & \makecell{29 \\ (43.94\%)} & \makecell{54 \\ (44.63\%)} & \makecell{34 \\ (48.57\%)} & \makecell{75 \\ (40.98\%)} & \makecell{88 \\ (46.07\%)} \\
    \textbf{3: The answer focuses on social activities on the platform (e.g., tagging, following, links sent by others) but not profiling or engagement.} & \makecell{7 \\ (5.98\%)} & \makecell{10 \\ (15.15\%)} & \makecell{7 \\ (5.79\%)} & \makecell{5 \\ (7.14\%)} & \makecell{17 \\ (9.29\%)} & \makecell{12 \\ (6.28\%)} \\
    \textbf{4: The answer refers to similar profiles, activities, or engagement patterns between Jarmo and other users.} & \makecell{0 \\ (0.00\%)} & \makecell{1 \\ (1.52\%)} & \makecell{9 \\ (7.44\%)} & \makecell{9 \\ (12.86\%)} & \makecell{1 \\ (0.55\%)} & \makecell{18 \\ (9.42\%)} \\
    \bottomrule
  \end{tabular}
  \label{s2}
\end{table}

\begin{table}[ht]
  \centering
   \caption{Number and proportion of students of each response per test and grade for the third open-ended question}
  \begin{tabular}{@{}
    >{\raggedright\arraybackslash}p{0.25\textwidth} 
    >{\raggedright\arraybackslash}p{0.10\textwidth} 
    >{\raggedright\arraybackslash}p{0.10\textwidth} 
    >{\raggedright\arraybackslash}p{0.10\textwidth} 
    >{\raggedright\arraybackslash}p{0.10\textwidth} 
    >{\raggedright\arraybackslash}p{0.10\textwidth} 
    >{\raggedright\arraybackslash}p{0.10\textwidth} @{}}
    \toprule
    & \textbf{Pre-test Grade 5} & \textbf{Pre-test Grade 8} & \textbf{Post-test Grade 5} & \textbf{Post-test Grade 8} & \textbf{Pre-test Total} & \textbf{Post-test Total} \\
    \midrule
    \textbf{0: No answer, "I don't know", or answers that do not address the question.} & \makecell{68 \\ (58.12\%)} & \makecell{37 \\ (56.06\%)} & \makecell{44 \\ (36.36\%)} & \makecell{24 \\ (34.29\%)} & \makecell{105 \\ (57.38\%)} & \makecell{68 \\ (35.6\%)} \\
    \textbf{1: The answer is limited to generic descriptions of good or addictive content without reference to targeting or engagement.} & \makecell{31 \\ (26.5\%)} & \makecell{11 \\ (16.67\%)} & \makecell{42 \\ (34.71\%)} & \makecell{12 \\ (17.14\%)} & \makecell{42 \\ (22.95\%)} & \makecell{54 \\ (28.27\%)} \\
    \textbf{2: The answer refers to targeting content to meet Jarmo's interests or liking, but does not mention how those interests or liking are determined (from data).} & \makecell{0 \\ (0.00\%)} & \makecell{1 \\ (1.52\%)} & \makecell{2 \\ (1.65\%)} & \makecell{2 \\ (2.86\%)} & \makecell{1 \\ (0.55\%)} & \makecell{4 \\ (2.09\%)} \\
    \textbf{3: The answer refers to targeting based on data collected from or inferred about Jarmo.} & \makecell{15 \\ (12.82\%)} & \makecell{15 \\ (22.73\%)} & \makecell{22 \\ (18.18\%)} & \makecell{12 \\ (17.14\%)} & \makecell{30 \\ (16.39\%)} & \makecell{34 \\ (17.8\%)} \\
    \textbf{4: The answer describes targeting based on engagement, emotional reactions, or profiling.} & \makecell{3 \\ (2.56\%)} & \makecell{2 \\ (3.03\%)} & \makecell{11 \\ (9.09\%)} & \makecell{20 \\ (28.57\%)} & \makecell{5 \\ (2.73\%)} & \makecell{31 \\ (16.23\%)} \\
    \bottomrule
  \end{tabular}
 
  \label{s3}
\end{table}

\begin{figure}
    \centering
    \includegraphics[width=\linewidth]{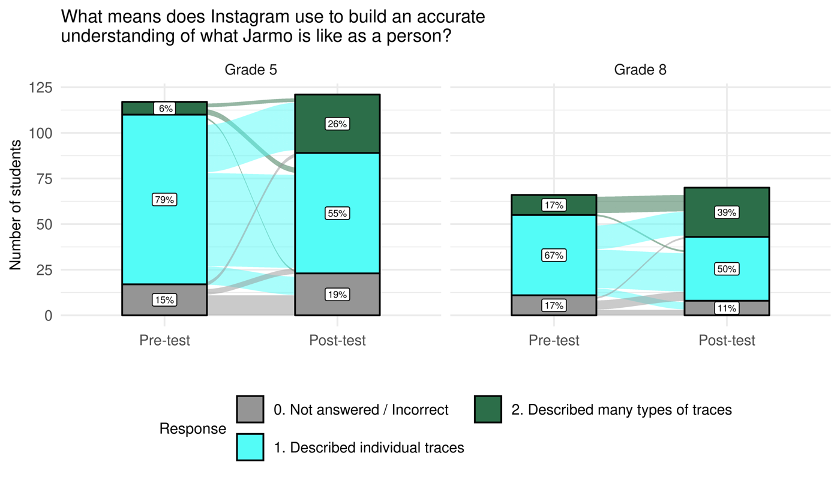}
    \caption{Alluvial plot of changes in students' responses between
pre-test and post-test per grade to the first open-ended question, which
assessed children's recognition of the variety of data traces.}
\label{fig:figs1}
\end{figure}

\begin{figure}
    \centering
    \includegraphics[width=\linewidth]{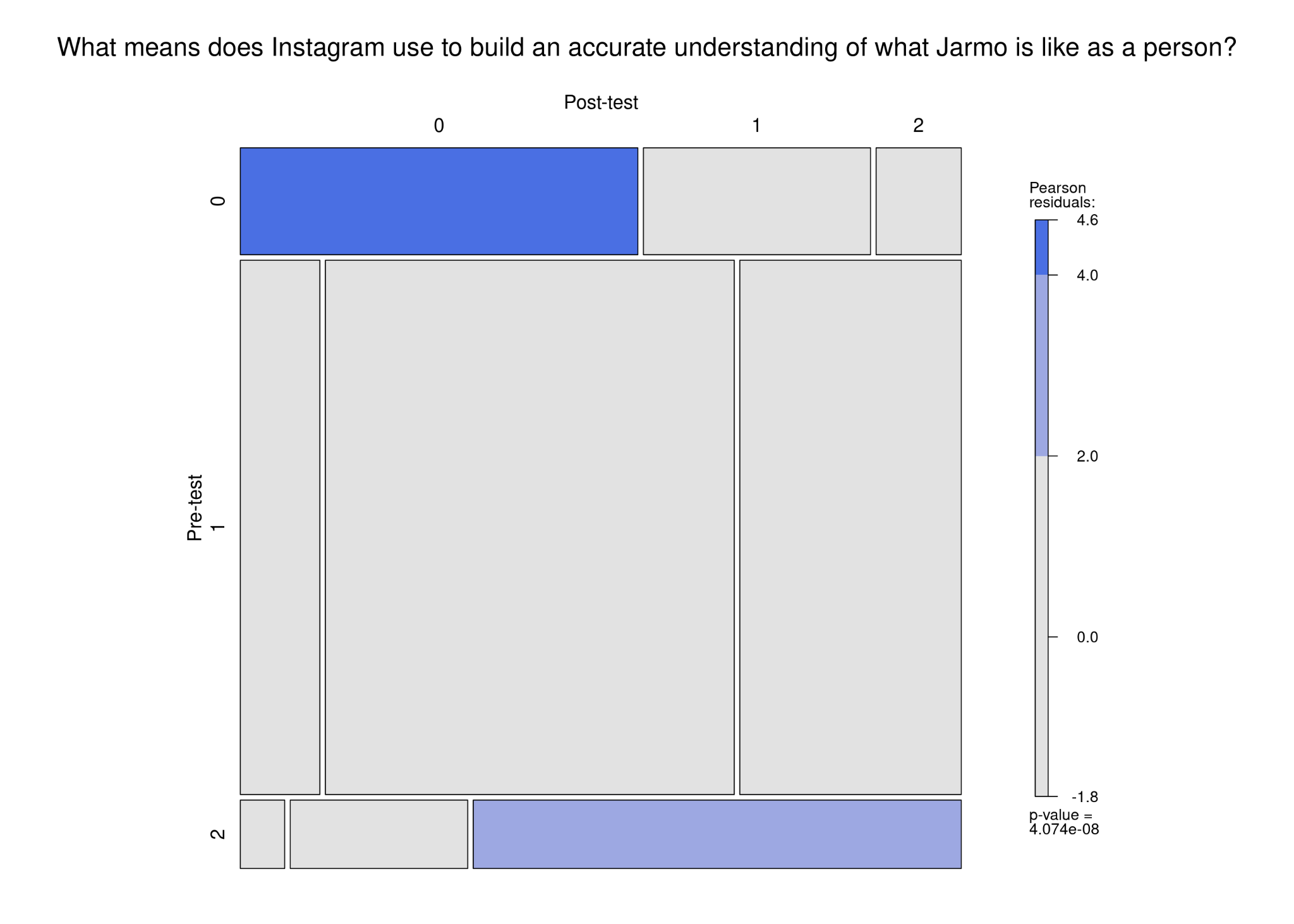}
    \caption{Mosaic plot representing the chi-squared test comparing
students' responses between pre-test and post-test to the first
open-ended question, which assessed children's recognition of the
variety of data traces. Cells are shaded according to Pearson's
residuals.}
\label{fig:figs2}
\end{figure}

\begin{figure}
    \centering
    \includegraphics[width=\linewidth]{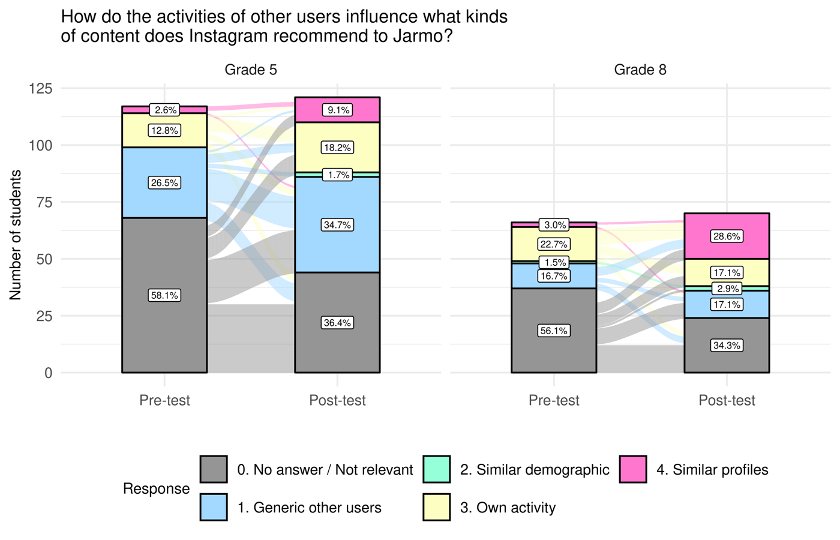}
    \caption{ Alluvial plot of changes in students' responses between
pre-test and post-test per grade to the second open-ended question,
which assessed children's awareness and understanding of social
profiling and recommending mechanisms on social media.}
\label{fig:figs3}
\end{figure}

\begin{figure}
    \centering
    \includegraphics[width=\linewidth]{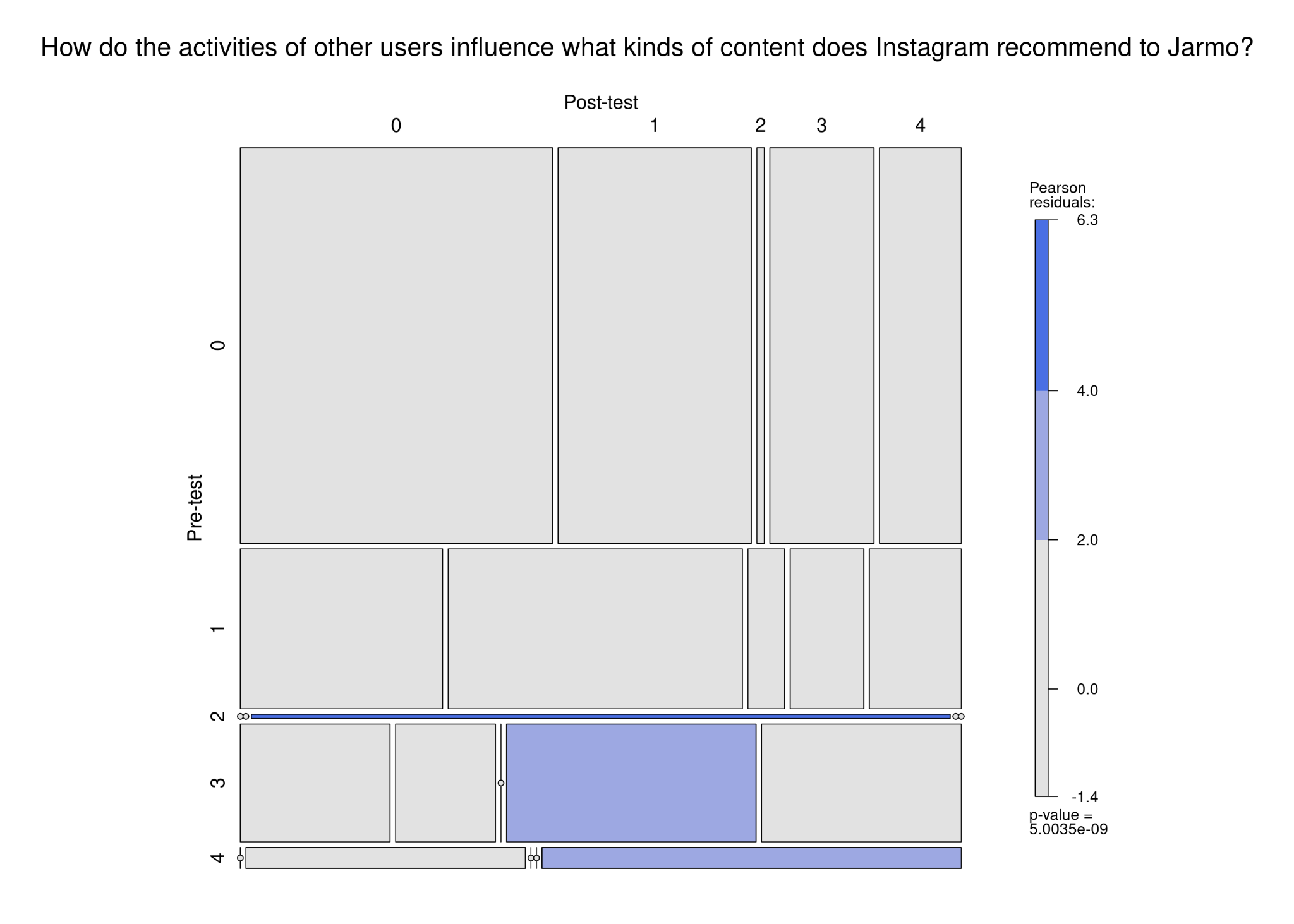}
    \caption{ Mosaic plot representing the chi-squared test comparing
students' responses between pre-test and post-test to the second
open-ended question, which assessed children's awareness and
understanding of social profiling and recommending mechanisms on social
media. Cells are shaded according to Pearson's residuals.}
\label{fig:figs4}
\end{figure}

\begin{figure}
    \centering
    \includegraphics[width=\linewidth]{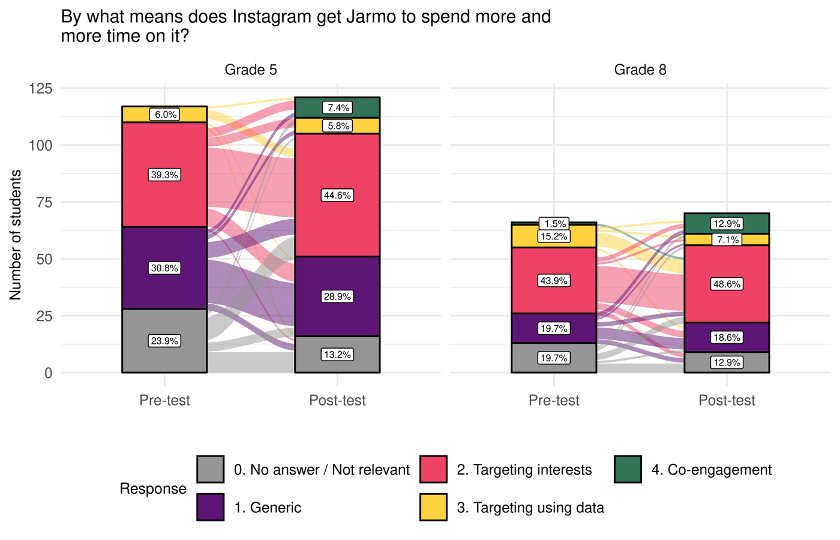}
    \caption{ Alluvial plot of changes in students' responses between
pre-test and post-test per grade to the third open-ended question, which
assessed children's awareness and understanding of engagement related
mechanisms on social media.}
\label{fig:figs5}
\end{figure}

\begin{figure}
    \centering
    \includegraphics[width=\linewidth]{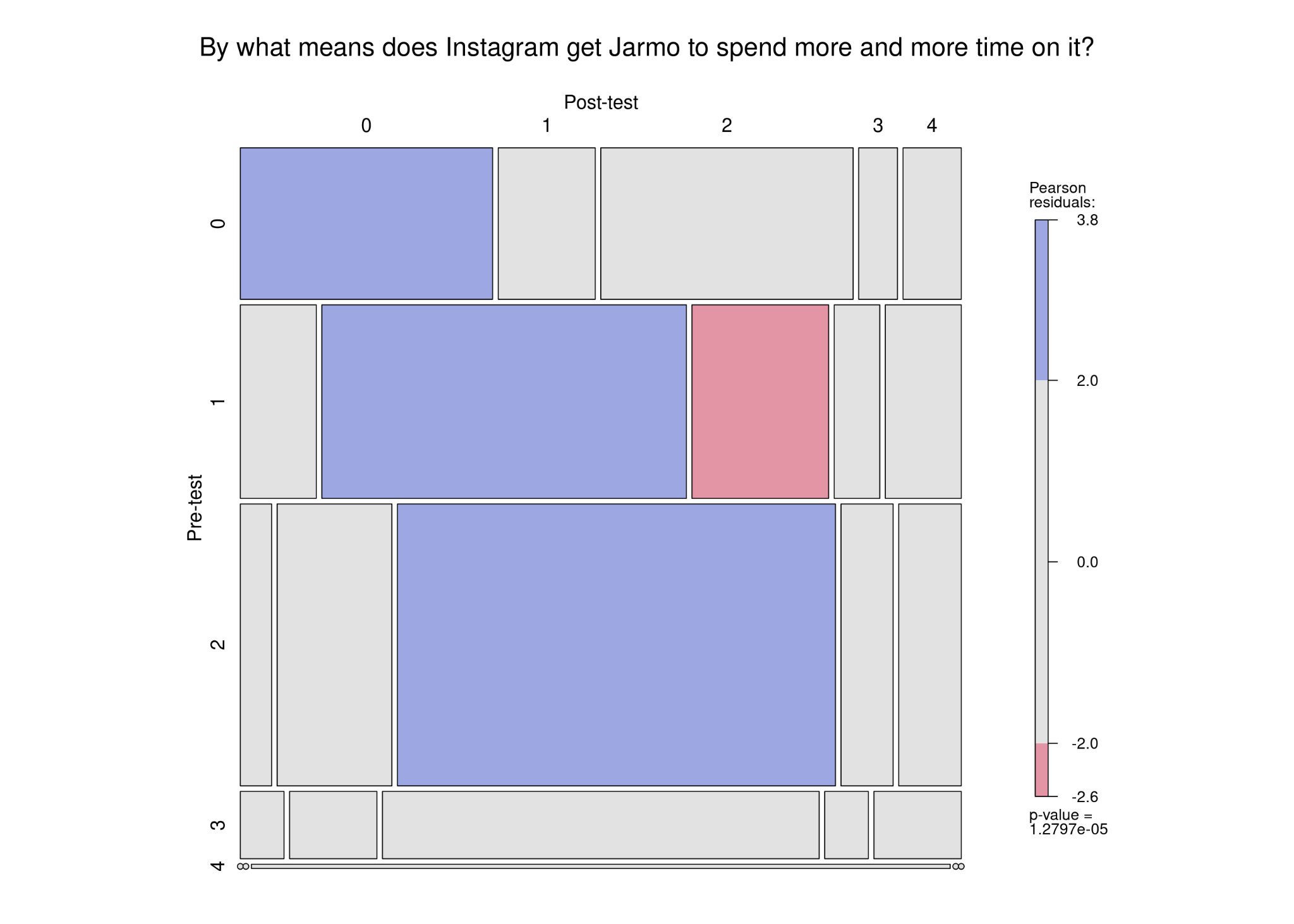}
    \caption{ Mosaic plot representing the chi-squared test comparing
students' responses between pre-test and post-test to the third
open-ended question, which assessed children's awareness and
understanding of engagement related mechanisms on social media. Cells are shaded according to Pearson's residuals.}
\label{fig:figs6}
\end{figure}

\end{document}